\documentclass[reprint,superscriptaddress]{revtex4-1}
\usepackage{graphics}
\usepackage{graphicx}
\usepackage{amsmath}
\usepackage{amssymb}
\usepackage{comment}
\usepackage{xcolor}

\begin{document}

\title{Universal behavior in fragmenting brittle, isotropic solids across material properties }

\author{Joel T. Clemmer}
\email{Correspondence to: jtclemm@sandia.gov}
\affiliation{Sandia National Laboratories, Albuquerque, New Mexico 87185, USA}
\author{Mark O. Robbins}
\thanks{Deceased.}
\affiliation{Department of Physics and Astronomy, Johns Hopkins University, Baltimore, Maryland 21218, USA}
\date{\today}

\begin{abstract}
A bonded particle model is used to explore how variations in the material properties of brittle, isotropic solids affect critical behavior in fragmentation.
To control material properties, a model is proposed which includes breakable two- and three-body particle interactions to calibrate elastic moduli and mode I and mode II fracture toughnesses. 
In the quasistatic limit, fragmentation leads to a power-law distribution of grain sizes which is truncated at a maximum grain mass that grows as a non-trivial power of system size.
In the high-rate limit, truncation occurs at a mass that decreases as a power of increasing rate. 
A scaling description is used to characterize this behavior by collapsing the mean-square grain mass across rates and system sizes.
Consistent scaling persists across all material properties studied although there are differences in the evolution of grain size distributions with strain as the initial number of grains at fracture and their subsequent rate of production depend on Poisson's ratio.
This evolving granular structure is found to induce a unique rheology where the ratio of the shear stress to pressure, an internal friction coefficient, decays approximately as the logarithm of increasing strain rate.
The stress ratio also decreases at all rates with increasing strain as fragmentation progresses and depends on elastic properties of the solid.
\end{abstract}

\maketitle

\section{Introduction}
\label{sec:intro}

The breakdown of brittle solids into smaller components, or comminution, is relevant to countless physical systems including, but not limited to, industrial processes such as milling, geomechanical fragmentation of rock, and ballistic impacts.
In such applications, it is of great importance that one can predict the state of the final fragmented product. 
To effectively design a milling process that optimizes a given powder property, such as the coarseness of flour for baking cookies vs. bread \cite{Gaines1985, Doblado-Maldonado2012}, one must understand the impact of numerous control parameters such as the speed of the mill and the total grinding time \cite{Canakci2013}. 
In geomechanics, the frictional stability of faults and the energy balance of earthquakes depend on the structure and breakup of rock \cite{Marone1989, Marone1990, Nguyen2009}.
During ballistic impacts, one may want to estimate the likelihood of fragmentation or predict debris sizes for asteroid collisions \cite{OKeefe1985, Ryan2000} particularly for planning redirection missions \cite{Stickle2022}.
Similar concerns are relevant to the design of ceramics for ballistic armoring \cite{Chen2007, Hogan2017, Ramesh2022}.
Therefore, there is a great need for continuum mechanical models of fragmentation in brittle materials \cite{Einav2007b, Einav2007c, Cil2019, Bhattacharjee2021}.

Beyond such applications, comminution is also theoretically interesting as a transformation from a solid to a complex granular state with dynamically evolving grain sizes and shapes.
Intriguingly, fragmentation has often been noted to result in a power-law distribution of grain masses $N(M) \sim M^{-\tau}$, where $N$ is the number of grains of a given mass $M$ and $\tau$ is an exponent with various measured values in different systems \cite{Turcotte1986}.
This has been identified in both experiments and simulations of different fragmented materials under various loading conditions \cite{Marone1989, Oddershede1993, Kun1996, Astrom1998, McDowell2002, Astrom2004, Coop2004, Carmona2008, Wittel2008, Timar2010, Altuhafi2011, Xiao2017, Xu2018, Iliev2019}. 

Due to these findings, comminution has been postulated to be an example of self-organized criticality \cite{Bak1988}, where the system is driven by shear, impact, or crushing to a critical state with a power-law distribution of grains \cite{Oddershede1993, Inaoka1996, Astrom1998, Einav2007}.
In recent work \cite{Clemmer2022}, we used large-scale particle-based simulations of fragmentation to characterize the development of the grain size distribution in brittle, isotropic solids under shear.
In the quasistatic, infinite-system-size limit, fragmentation produced a power-law distribution of grains.
Moving away from this limit by introducing finite-size or finite-rate effects, this power law was truncated by a power of either decreasing system size or increasing strain rate with nontrivial exponents.
In analogy to critical scaling theories for the magnitudes of avalanches in the depinning and yielding transitions \cite{Cote1991, Ji1992, Fisher1998, Perkovic1999, Sethna2001, Sun2010, Hayman2011, Salerno2012, Salerno2013, Lin2014, Bares2017, Moura2017, Clemmer2019}, we proposed a scaling theory for the distribution $N(M)$ and measured exponents by collapsing grain size distributions and their moments across rates and system sizes.

An important follow-up question is whether these results depend on the specific model or material being fragmented or whether they demonstrate universal behavior.
While fundamental changes in the class of material, e.g., brittle vs. plastic materials \cite{Timar2010}, may be associated with significant changes in critical behavior, does the same scaling theory describe all isotropic, brittle solids such that critical exponents are insensitive to the material properties of the system?
Furthermore, if the critical behavior is universal, are there other important non-critical aspects of fragmentation that depend on the material properties?

Experimental studies of impacted objects found evidence suggesting the power-law exponent $\tau$ may not depend on the material being fractured \cite{Oddershede1993}.
Similar values of $\tau$ have also been found in various other simulations and experiments of brittle impacts \cite{Astrom2000, Astrom2004, Carmona2008, Timar2010}.
In this work, we alternatively consider a shear geometry which provides continuous fragmentation.
While most experiments consider either impact \cite{Oddershede1993, Timar2010} or compaction \cite{Altuhafi2011, Xiao2017}, some experiments use shear geometries and find a power-law grain size distribution that continues to evolve until large strains \cite{Marone1989, Coop2004}.
We also extend upon investigations of universality in fragmentation by systematically controlling material properties to limit uncertainty and by characterizing how other aspects of fragmentation depend on material properties.
To accomplish this task, we build upon the bonded particle model in the previous work which only included two-body bond interactions by adding three-body angular interactions. 
With this modification, the model has three parameters which can control Poisson's ratio and the mode I and mode II fracture toughnesses.

With variations in material properties, we identify changes in noncritical behavior such as the onset of power-law scaling in the grain size distribution.
However, no significant deviation from the scaling theory or exponents is detected with the exception of a possible exponent that may govern the growing number of grains with strain.
In particular, we find there are initially fewer grains at fracture at higher Poisson's ratios but the number of grains subsequently grows at a faster rate with increasing strain relative to systems with small Poisson's ratios.
While we primarily focus on the role of elasticity in the fragmentation of two-dimensional systems in this work, we also briefly explore changes in the fracture toughness and highlight key results for three-dimensional systems.

In addition, we expand upon our previous work and characterize the rheology of the system.
In contrast to typical studies of granular flow, the grain size distribution in comminuting systems evolves with strain.
We find this leads to a reduction in the ratio of the shear stress to pressure, or the internal frictional resistance to flow, with increasing strain.
Furthermore, we also identify that the grain size distribution is strongly dependent on the strain rate, finer grains are produced at higher rates, which produces a unique rheology where the stress ratio decreases approximately logarithmically with increasing rate.
This behavior provides useful insight into the logarithmic weakening described by theories of rate and state friction \cite{Daub2010}.

The remainder of this article is organized as follows.
First we present the methodology and describe the numerical model, its calibration, and the deformation protocol of simulations in Secs. \ref{sec:model}, \ref{sec:calibration}, and \ref{sec:deformation}, respectively.
Next, we discuss results starting with a description of how the system fractures and fragments with strain at different Poisson's ratios in Sec. \ref{sec:strain}.
Finite-size, finite-rate, and combined finite-size and finite-rate effects are presented in Secs. \ref{sec:size}, \ref{sec:fr}, and \ref{sec:frs}, respectively, along with a scaling theory for the grain size distribution.
The impact of fracture toughness is discussed in Sec. \ref{sec:kc} and rheology is discussed in Sec. \ref{sec:rheo}
The results of this article are summarized in Sec. \ref{sec:summary}.

\section{Methodology}
\label{sec:methods}

Fragmentation involves many physical mechanisms which can be challenging to simulate, including elastic deformation, contact forces, and crack nucleation, growth, and coalescence.
Beyond the need to represent all of these mechanisms, a model must also be computationally efficient and scale to large system sizes to resolve highly polydisperse systems with a representative number of grains.
While there are mesh-based continuum techniques which can model fracture \cite{Rabczuk2013,Rege2017}, particle-based models are particularly well suited for this problem as they naturally handle the large number of discontinuities present in the fragmentation and flow of granular materials.
In this article, the term particle is only used to refer to the fundamental element of a simulation which may not correspond to a physical grain.
Grains are composed of one or more particles.

Among particle-based models, there are many continuum mesh-free methods that can model cracking \cite{Sun2021} and several have already been applied to the fragmentation of granular materials including the material point method \cite{Homel2017} and peridynamics \cite{Behzadinasab2018}. 
These approaches are ideal when one needs to model a specific material with a given constitutive equation relating stress and strain.
Another particle-based technique, which is particularly popular for modeling comminution, is the discrete-element method (DEM) \cite{Cundall1979}.
In DEMs, particles traditionally each represent an individual grain and interact by exchanging pairwise forces and torques while numerically integrating their translational and rotational degrees of freedom using Newtonian mechanics.
While the DEM is generally quite computationally efficient, it can be a challenge to design and calibrate interactions to reproduce the behavior of specific materials \cite{Windows-Yule2022}.

To introduce fracture to DEMs, several ideas have been proposed.
One approach is to ignore the dynamics of crack growth and replace a large particle with multiple smaller particles when the stress on the large particle exceeds some critical threshold to induce fracture. 
This can be done by either splitting a grain along predicted crack paths as in level-set DEM \cite{Harmon2020} or replacing particles with a predefined collection of smaller particles \cite{Astrom1998, Ben-Nun2010}.
While well-designed splitting rules may produce realistic results \cite{Jiang2020}, these techniques are fundamentally limited to slow strain rates as they assume a separation in timescales between the external loading of a grain and the dynamics of fracture.
Another approach is to represent a grain using a collection of particles linked by a network of attractive bonds \cite{Herrmann1989, Mora1994, Jirasek1995, Potyondy2004, Carmona2008, Wang2008b, Bobet2009, Andre2012, Lisjak2014}.
Fracture is modeled by allowing bonds to break.
There are many variations of these models with many different names, so for simplicity we refer to them using an umbrella term of bonded-particle models (BPMs). 

For this work we designed a minimal BPM based on an early model by Maloney and Robbins \cite{Maloney2007} revised in more recent work \cite{Clemmer2022} which uses point particles, akin to a breakable spring network \cite{Beale1988, Hassold1989}.
Unlike typical DEM-like BPMs, we do not resolve the rotational degrees of freedom of particles as rotation still emerges in collections of bonded particles.
This reduces the computational complexity of evaluating forces and integrating trajectories allowing us to simulate larger systems and longer run times, needed to probe critical behavior.
Bonds have equilibrium lengths equal to their initial length, creating a stress-free reference state, and break if stretched in tension beyond a threshold.
This approach also provides substantial flexibility to implement a diverse range of mechanical responses such as plastic deformation \cite{Clemmer2023} with minimal changes to the model.
In this article we introduce three-body interactions which expands control over material properties, a feature of the Kirkwood-Keating spring model \cite{Kirkwood1939, Keating1966}.
Angular terms have similarly been use in studies of pruning disordered spring networks \cite{Reid2018}.
The model was implemented in the Large Scale Atomic/Molecular Massively Parallel Simulator (LAMMPS) \cite{Plimpton1995, Thompson2021}, which now has a dedicated BPM package.

\subsection{Model details}
\label{sec:model}

In three dimensions (3D), particles are monodisperse with diameters of $a$, the unit of length.
To prevent crystallization in 2D, particles are bidisperse with diameters of $3/5a$ and $a$.
The ratio of the number of large particles to small particles is $(1+\sqrt{5})/4$ as in Ref. \cite{Maloney2006}.
All particles have mass $m$.
Disordered packings of particles with periodic boundary conditions are generated using an initialization protocol similar to that in Refs. \cite{Maloney2006, Maloney2008, Clemmer2021a, Clemmer2021b}.
Using a disordered packing ensures solids are isotropically elastic and avoids anisotropic crack growth associated with regular lattices \cite{Jirasek1995, Zhou1996, Gumbsch1997}. 

\begin{figure}
\begin{centering}
	\includegraphics[width=0.45\textwidth]{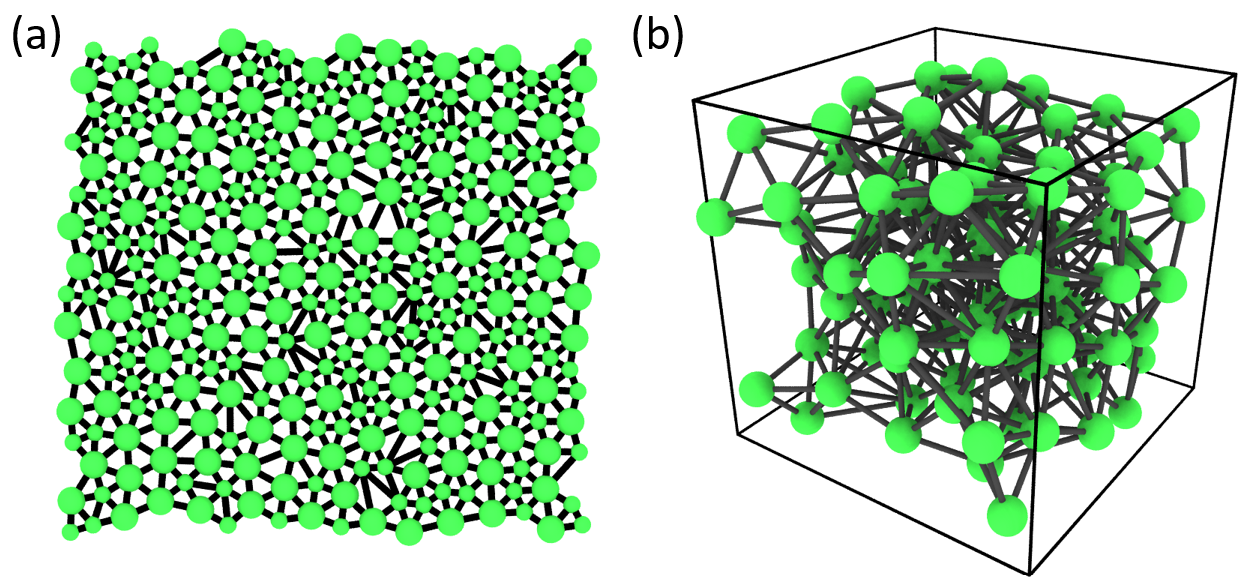}
	\caption{Example starting states in (a) 2D and (b) 3D. Bonds that cross a periodic boundary are not rendered.} 
	\label{fig:example_system}
\end{centering}
\end{figure}

Four types of interactions are used in simulations: a pairwise non-bonded repulsion $F_\mathrm{NB}$, a pairwise bonded force $F_\mathrm{B}$, a three-body angular force $F_\mathrm{A}$, and a pairwise damping force $F_\mathrm{D}$.
These interactions are formulated to use a minimal number of free parameters while also conserving linear and angular momentum and avoiding discontinuous forces.
A list of bonds and angles are generated using a Delaunay triangulation of the initial particle packing.
In 3D, each edge with a length less than $5/4 \times 2^{1/6} a$ is used to generate a bond.
This maximum length criterion did not have any significant impact on results but reduced the necessary communication between processors for simulations running in parallel.
For each triangle in 2D or triangular face of a tetrahedron in 3D, a three-body angular bond is created at every vertex unless one of its associated bonds were pruned due to the length restriction.
Each particle is associated with 6 pairwise bonds and 6 angular bonds in 2D and $\sim 11$ pairwise bonds and $\sim 24$ angular bonds on average in 3D, well above the Maxwell rigidity criterion \cite{Liu2010}.
Examples of small starting states representing bulk, unfractured material are rendered in Fig. \ref{fig:example_system}

The central-body, non-bonded interaction represents contact forces between particles on opposite sides of a crack or particles from separate grains that are within a distance $r$ less than the sum of particle radii $\bar{a}$. 
Its magnitude is calculated using a truncated, purely repulsive Lennard-Jones force with a minimum at $\bar{a}$,
\begin{equation}
F_{NB} = 
\begin{cases}
  \frac{12 u}{r} \left( \frac{\bar{a}^{12}}{r^{12}} - \frac{\bar{a}^{6}}{r^{6}} \right) , & r < \bar{a} \\
  0 , & r > \bar{a}
\end{cases} 
\end{equation}
where $u$ is the unit of energy. 
Bond forces represent elastic interactions within a solid body and are central body with a magnitude that also depends on the initial distance $r_0$ between the particles,
\begin{equation}
F_{B} = 
\begin{cases}
  \frac{6 \times 2^{2/3} u r_0^2 }{a^2 r} \left( \frac{r_0^{12}}{r^{12}} - \frac{r_0^6}{r^6} \right) , & r < r_0 \\
C_1 (r_0 - r) + C_3 (r_0 - r)^3, & r_0 < r < \lambda_c r_0 \\
0, & r > \lambda_c r_0    
\end{cases}
\end{equation}
where $\lambda_c$ represents the maximum stretch $r/r_0$ of the bond and the $C$ coefficients are $C_1 = 36 \times 2^{2/3} u a^{-2}$ and $C_3 = -36 \times 2^{2/3} (\lambda_c - 1)^{-2} u r_0^{-2} a^{-2}$.
These coefficients are chosen to ensure that all bonds have a constant $r_0$-independent linear stiffness of $k_{B} = 36 \times 2^{2/3} u a^{-2}$ around an equilibrium distance of $r = r_0$ and that forces go to zero at $r = \lambda_c r_0$, the limit where bonds break. 

Every three-body angular interaction is associated with two bonds that share a central particle.
This interaction is a function of the deviation between the initial and current angles between the two bonds, $\delta \theta \equiv \theta - \theta_0$, as well as the current stretch of the two bonds, $\lambda_1$ and $\lambda_2$.
Three-body forces act as torque springs within the plane of the three particles with a magnitude of
\begin{equation}
F_{A} =
k_A S(\lambda_1, \lambda_2) \times \begin{cases}
\delta \theta - \frac{1}{\theta_c^2}  \delta \theta^3, & |\delta \theta| < \theta_c \\
0, & |\delta \theta| > \theta_c
\end{cases}
\end{equation}
where $k_A$ is an angular stiffness, $\theta_c$ represents the maximum angular deviation from $\theta_0$, and $S(\lambda_1, \lambda_2)$ is a smoothing term given by
\begin{equation}
S(\lambda_1, \lambda_2) = 
\begin{cases}
1, & \epsilon_\mathrm{max} < 0 \\
1 - 2  \frac{\epsilon_\mathrm{max}^2}{\epsilon_c^2} + \frac{\epsilon_\mathrm{max}^4}{\epsilon_c^4}, & 0 < \epsilon_\mathrm{max} < \epsilon_c \\
0, & \epsilon_\mathrm{max} > \epsilon_c
\end{cases}
\end{equation}
where $\epsilon_c \equiv \lambda_c - 1$ and $\epsilon_\mathrm{max} \equiv \mathrm{max}(\lambda_1, \lambda_2)-1$.
Forces smoothly go to zero and the angular interaction breaks as $\delta \theta$ approaches $\theta_c$ or as either of the associated bonds breaks.

Finally, a damping force, commonly used in dissipative particle dynamics \cite{Groot1997}, is also applied to all pairwise interacting particles,
\begin{equation}
\vec{F}_D = - \gamma \left(1 -\frac{r}{r_\mathrm{max}} \right)^2 (\hat{r} \cdot \delta\vec{v}) \hat{r}
\end{equation}
where $\vec{r} = r \hat{r}$ is the vector between the two particle positions, $\delta \vec{v}$ is the difference in particle velocities, $r_\mathrm{max}$ represents the maximum interaction distance ($\bar{a}$ for non-bonded particles and $\lambda_c r_0$ for bonded particles), and $\gamma$ is the damping strength.
This construction is Galilean invariant and is the lowest-order damping term present in isotropic solids \cite{Sethna2006}. 
The damping strength $\gamma$ does not significantly affect material properties and is set to $50 \sqrt{m u}/a$ in all simulations, large enough to ensure there are minimal thermal effects.
For the remainder of this article, all quantities are reported without units, scaled by the necessary factors of $a$, $m$, and $u$.
A velocity-Verlet integrator is used with a time step of 0.005. 

\subsection{Calibration of material properties}
\label{sec:calibration}

\begin{figure}
\begin{centering}
	\includegraphics[width=0.40\textwidth]{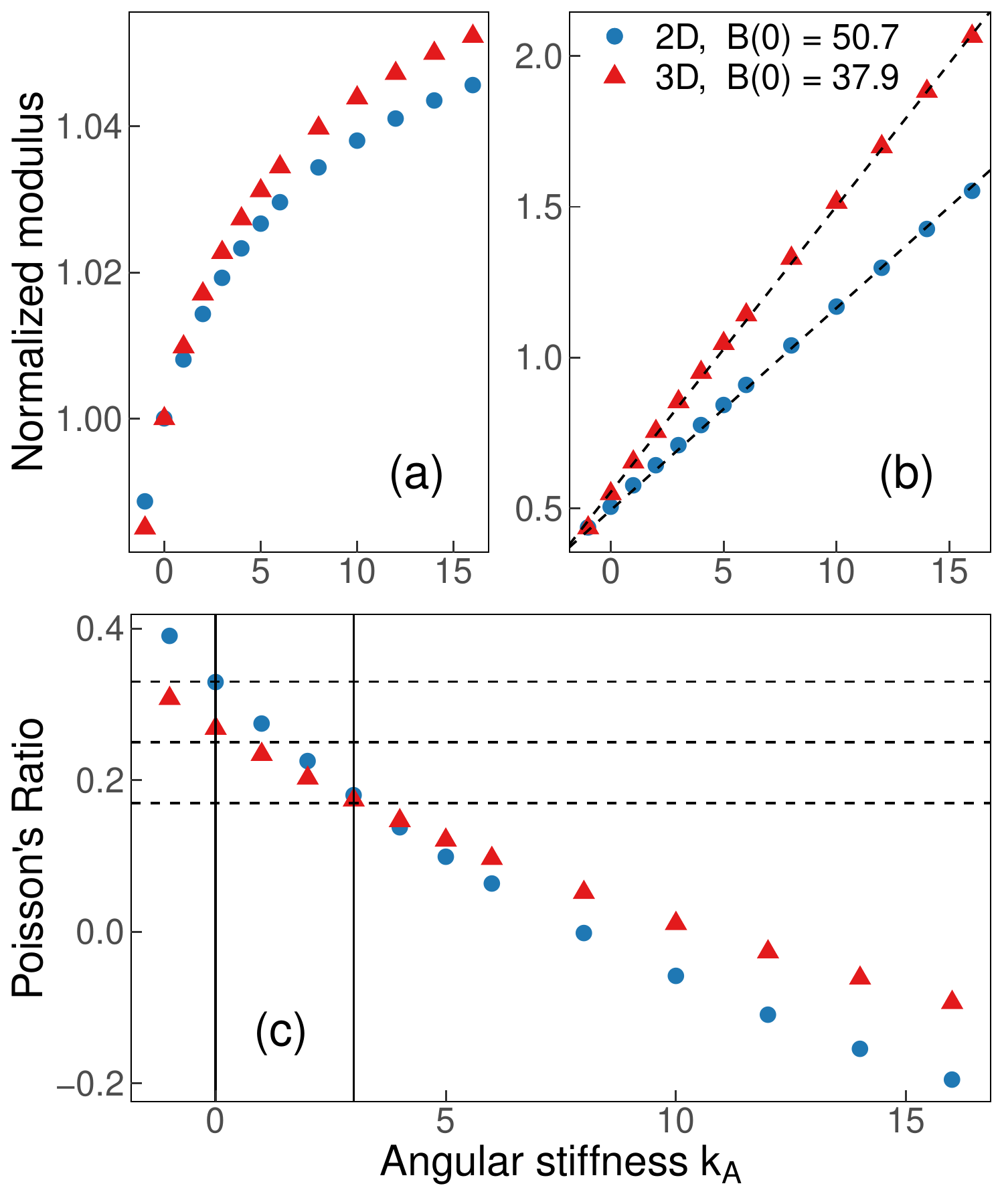}
	\caption{(a) Bulk and (b) shear modulus normalized by the bulk modulus at $k_A = 0$, or $B(0) = 50.7$ in 2D and $37.9$ in 3D, as a function of angular stiffness $k_A$ in 2D (blue circles) and 3D (red triangles). Dashed lines in (b) have slopes of 3.4 and 3.6 and intercepts of 25 and 21 in 2D and 3D, respectively. (c) Poisson's ratio $\nu_\mathrm{PR}$ as a function of $k_A$. Horizontal dashed lines highlight $\nu_\mathrm{PR} = 0.17$ (the approximate value of boron carbide), 1/4, and 1/3. Vertical lines mark $k_A = 0$ and 3.0.} 
	\label{fig:moduli}
\end{centering}
\end{figure}

After accounting for units, the model described above has three free parameters:  $k_A$, $\lambda_c$, and $\theta_c$, i.e., an angular stiffness, critical bond stretch, and critical angle.
In this section, these three parameters are mapped to three important material properties and calibrated.
For isotropic systems, linear elasticity can be described by two variables such as the bulk and shear moduli.
As $\lambda_c$ and $\theta_c$ affect failure at large strains, the only relevant parameter is $k_A$.
The bulk modulus $B$ is measured by isotropically compressing a periodic sample at a constant strain rate and fitting the resulting linear rise in pressure up to a volumetric strain of $0.5\%$. 
The response to volumetric expansion or contraction is largely determined by pairwise bonds and $B$ has minimal dependence on $k_A$ (Fig. \ref{fig:moduli}[a]).
We therefore assume a fixed value of $50.7$ in 2D and $37.9$ in 3D.
This value can be scaled to match the bulk modulus of real materials by simply adjusting the units of a simulation.
The shear modulus $G$ is similarly measured using simple shear deformation and is found to linearly increase with $k_A$ as seen in Fig. \ref{fig:moduli}(b). 
From these results, the Poisson's ratio, $\nu_\mathrm{PR} = (B-G)/(B+G)$ in 2D and $(3B-2G)/(6B+2G)$ in 3D, depends significantly on $k_A$ in Fig. \ref{fig:moduli}(c). 

In isotropic particle packings with pairwise central forces, Poisson's ratio is restricted to equal a third in 2D and a quarter in 3D \cite{Walton1987}, as approximately obtained above in the absence of angular forces at $k_A = 0$.
By increasing the strength of angular interactions, triangular or tetrahedral collections of bonded particles become more resistant to changes in shape.
In isotropic compression or expansion, this is largely irrelevant explaining why $B$ has minimal dependence on $k_A$. 
In contrast, this strengthens the material's shear modulus thus decreasing its Poisson's ratio, extending into the auxetic limit.

A similar effect is obtained in BPMs based on the DEM where particles have rotational degrees of freedom and bonds transmit shear forces and torques, acting like elastic beams.
By increasing the relative shear stiffness, these models can also represent smaller Poisson's ratios in disordered systems by similarly increasing the shear strength \cite{Wang2010,Andre2012}.
However, it is difficult to obtain the opposite result and decrease $G$ or increase Poisson's ratio.
One could potentially use negative shear or angular stiffnesses, as seen in Fig. \ref{fig:moduli}(c), but this may create instabilities.
Other unique solutions to control Poisson's ratio also exist.
In the Distinct Lattice Spring Model \cite{Zhao2011}, a local strain tensor is calculated at the location of each point particle to construct rotationally invariant shear springs.
In the Lattice Particle Model, a volumetric energy term is used to derive a multibody interaction between particles \cite{Chen2014}.

\begin{figure}
\begin{centering}
	\includegraphics[width=0.40\textwidth]{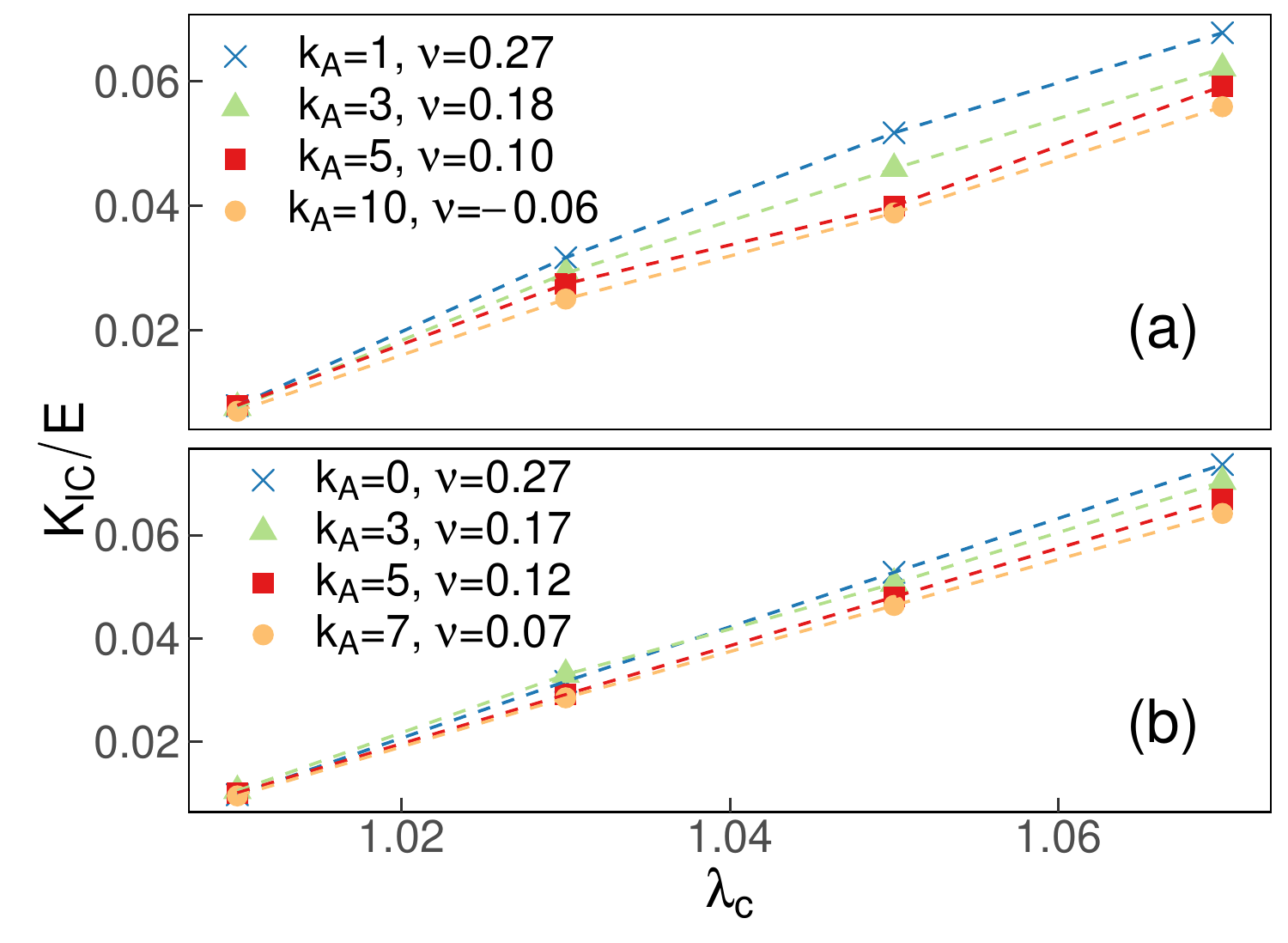}
	\caption{Mode I fracture toughness $K_\mathrm{IC}$ normalized by Young's modulus $E$ as a function of the critical bond stretch $\lambda_c$ for values of the angular stiffness $k_A$ and the approximate associated Poisson's ratios indicated in the legend in (a) 2D and (b) 3D.} 
	\label{fig:k1c}
 \end{centering}
\end{figure}

Next we consider the fracture toughness of the material, or the maximum stress intensity factor a crack in the solid can support before it propagates. 
We focus specifically on mode I and mode II fracture toughnesses which measure the resistance to the propagation of a tensile opening crack (mode I) and a shear crack (mode II).
To measure mode I toughness or $K_\mathrm{IC}$, we consider a square system with an elliptic void with major and minor axes of 20 and 2, respectively, and free boundaries.
In 3D, the third dimension is periodic and thin. 
All bonds crossing the ellipse are deleted. 
Tension is applied perpendicular to the major axis of the ellipse by displacing particles on the boundaries at a fixed rate until the crack grows and the system fails.
The peak stress before crack propagation is used to calculate $K_\mathrm{IC}$ \cite{Tada2000}. 
Varying the system size, crack length or width, and box height can change estimates of $K_\mathrm{IC}$ by about $10\%$.
As $\lambda_c$ sets the failure strain, $K_\mathrm{IC}$ increases linearly with $\lambda_c$ as seen in Fig. \ref{fig:k1c} where $K_\mathrm{IC}$ is normalized by Young's modulus $E$ to remove most of the dependence on $k_A$.
For small values of $\lambda_c$, no significant dependence of $K_\mathrm{IC}$ on $\theta_c$ is detected.
However at large values of $\lambda_c$, $K_\mathrm{IC}$ can depend up to $10\%$ on $\theta_c$ as angular bonds break before pairwise bonds. 
As this effect is relatively small, it is neglected.
The speed of crack growth is also measured by tracking the location of broken bonds in time. 
In 3D, the crack front accelerates at small times before reaching a constant speed of approximately 55\% of the Rayleigh wave speed, a theoretical maximum limit \cite{Washabaugh1994}.

\begin{figure}
\begin{centering}
	\includegraphics[width=0.35\textwidth]{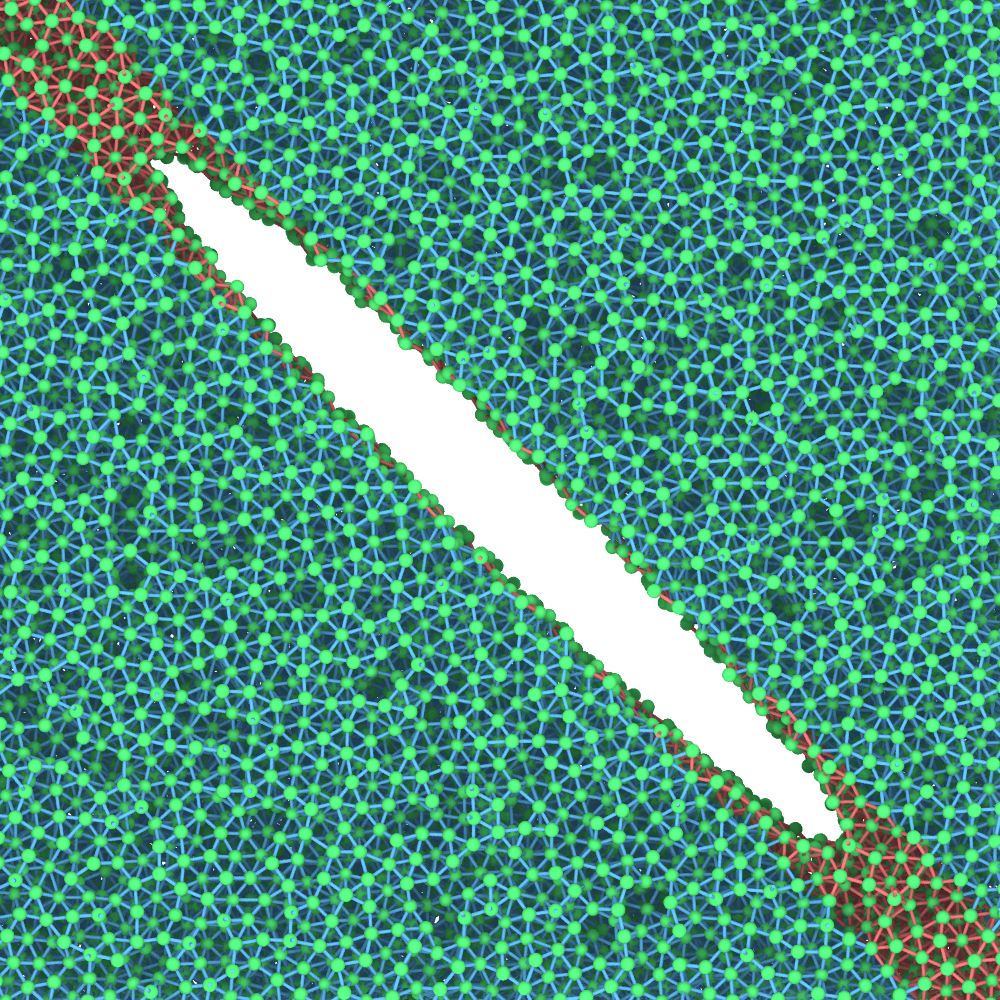}
	\caption{Geometry used to induce a shear crack in 3D. Compression is applied in the horizontal dimension while extension is applied in the vertical dimension. Only the red bonds have a finite $\lambda_c$.} 
	\label{fig:param_constrain_geom}
\end{centering}
\end{figure}

\begin{figure}
\begin{centering}
	\includegraphics[width=0.40\textwidth]{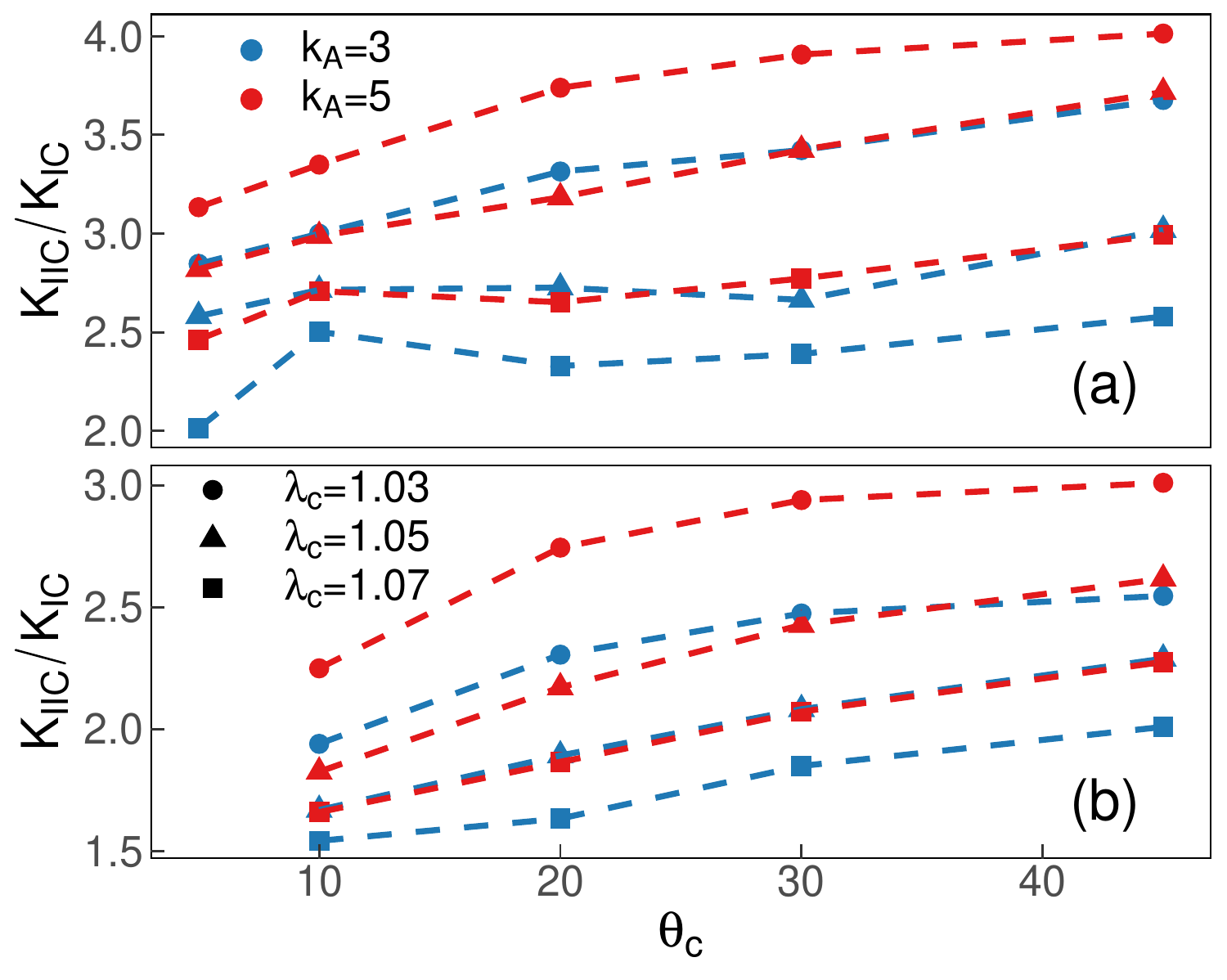}
	\caption{Ratio of mode II fracture toughness $K_\mathrm{IIC}$ to mode I fracture toughness $K_\mathrm{IC}$ as a function of the critical angle $\theta_c$ in (a) 2D and (b) 3D for $\lambda_c = 1.03$ (circles), 1.05 (triangles), and 1.07 (squares) and $k_A = 3$ (blue) and 5 (red). At $k_A = 3$, the Poisson's ratio is $\nu_\mathrm{PR} \approx 0.18$ in 2D and $0.17$ in 3D.} 
	\label{fig:k2c}
\end{centering}
\end{figure}

Finally, we measure the resistance to shear crack growth, or mode II fracture toughness $K_\mathrm{IIC}$. 
Experimentally inducing a pure mode II crack requires complex setups to suppress the growth of mode I cracks \cite{Backers2002, Rao2003}.
Here we force shear crack growth by controlling which bonds break in a simple geometry to estimate $K_\mathrm{IIC}$.
A fully periodic system undergoes pure shear with an elliptic void oriented $45^\circ$ between the tensile and compressive dimensions.
In 3D, the elliptic void extends through a thin third periodic dimension.
Only bonds along a thin region oriented in the direction of the crack, as rendered in Fig. \ref{fig:param_constrain_geom}, have a finite value of $\lambda_c$ and can break.
Then $K_\mathrm{IIC}$ is calculated from the peak shear stress \cite{Tada2000}.
The ratio of fracture toughnesses has a complicated dependence on $\lambda_c$, $k_A$, and $\theta_c$, as seen in Fig. \ref{fig:k2c}.
Generally there is an increase in $K_\mathrm{IIC}/K_\mathrm{IC}$ with increasing $\theta_c$, demonstrating that it is possible to independently vary the two fracture toughnesses.
However, we do not further explore this calibration as there is relatively limited experimental data on $K_\mathrm{IIC}$ in brittle materials for comparison.

\subsection{Simulations and deformation protocol}
\label{sec:deformation}

\begin{figure*}
\begin{centering}
	\includegraphics[width=0.98\textwidth]{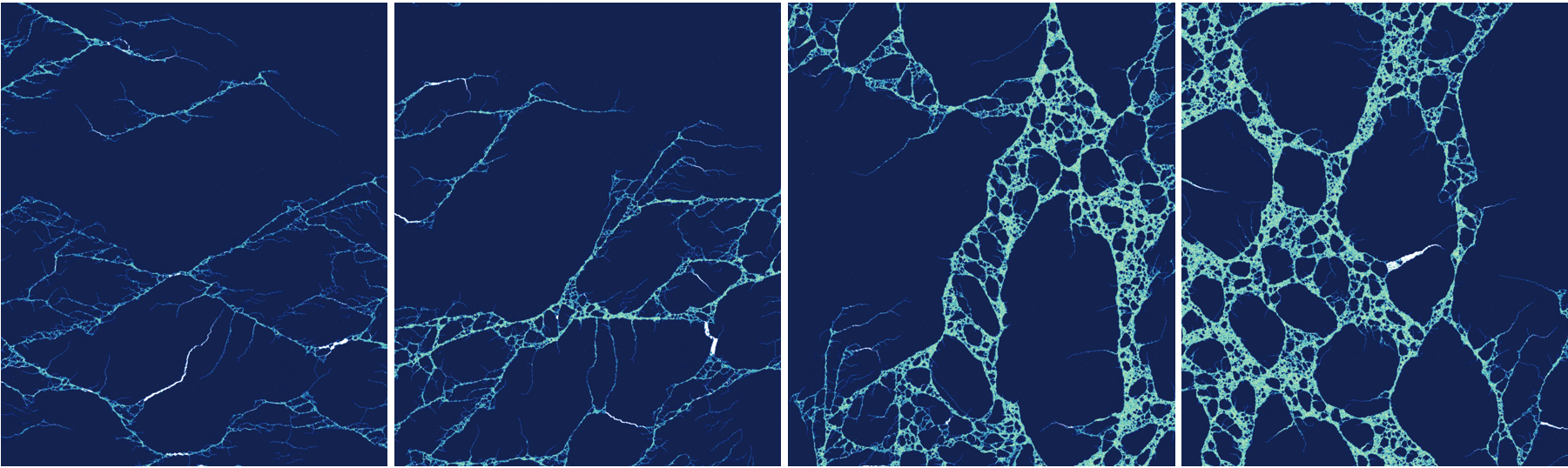}
	\caption{Sections of a 2D system with $k_a = 2.5$ sheared at a rate of $\dot{\epsilon} = 10^{-5}$ to strains of 0.02, 0.5, 1.0, and 2.0 going from left to right (or 2\% to 200\%). Particles are colored by the number of broken bonds going from zero (dark blue) to six (light yellow). White regions represent voids, such as gaps in opening cracks. The inset in the first panel highlights broken fragments along a crack.} 
	\label{fig:render_strain}
\end{centering}
\end{figure*}

Simulations of fragmentation are run using square or cubic systems with fully periodic boundaries and side lengths $L$ of $1600$ and $200$ with $~\sim 4 \times 10^6$ and $\sim 7 \times 10^6$ particles in 2D and 3D, respectively.
Section \ref{sec:size} also considers smaller systems in 2D.
As a reminder, these lengths as well as all others in the text are normalized by the diameter of particles $a$.
For each system size, multiple realizations are generated using different random packings to improve statistics by averaging results across these realizations. 
With the exception of Sec. \ref{sec:kc} which explores the effect of fracture toughness on fragmentation, simulations use values of $\lambda_c = 1.05$ and $\theta_c = 10^\circ$.
In contrast, many different values of $k_a$ are explored throughout the results.

After generating fully bonded initial states, constant volume shear is applied at a fixed strain rate denoted by$\dot{\epsilon}$.
In 2D, pure shear is applied by extending the $x$-dimension at a constant true strain rate of $\dot{\epsilon}$ while compressing the $y$-dimension by $-\dot{\epsilon}$.
In 3D, triaxial compression is applied by expanding the $x$- and $y$-dimensions at a rate $\dot{\epsilon}/2$ while compressing the $z$-dimension at a rate of $-\dot{\epsilon}$.
The strain is then defined as $\epsilon \equiv \dot{\epsilon} t$ where $t$ is the elapsed time.
Although this work considers only triaxial compression, it is important to note that in 3D there is a spectrum of shear types which can have significant impacts on granular flow \cite{Clemmer2021c} and could have impacts on fracture and fragmentation.
As the simulation cell deforms, particle positions are affinely remapped.
To avoid compressive dimensions becoming too thin, Kraynik-Reinelt boundaries are used in 2D \cite{Kraynik1992} and generalized Kraynik-Reinelt boundaries are used in 3D \cite{Dobson2014, Hunt2016}, leveraging various implementations in LAMMPS \cite{Nicholson2016, Clemmer2021a, Clemmer2021b}.
A stress tensor is calculated using the sum of the virial and kinetic energy tensors.

\section{Results and Discussion}
\label{sec:results}

A wide range of simulations were run to explore the impact of strain, strain rate, system size, and material properties on fragmentation.
Due to the extra computational costs incurred by the three-body angular interactions and the larger parameter space, results in this article are based on smaller system sizes and fewer random realizations than our previous work \cite{Clemmer2022}.
Therefore, we rely on our previous measures of exponents taken from systems with $k_a = 0$, summarized in Table \ref{table:exponents}, and generally do not attempt to refine estimates of exponents.
Instead, we primarily seek to identify where behavior changes with material properties and explore new results such as the rheology.

\begin{table}
\centering
\begin{tabular}{ |c|c|c|c|}
 \hline 
 Exponent & Estimate & Definition \\
 \hline
  $\ \ \tau\ \ $   & $\ 1.70 \pm 0.08 \ $  & $\ N(M) \sim M^{-\tau} \ $\\
  $\phi$           & $0.55 \pm 0.07$      & $N(M) \sim \epsilon^{\phi}$\\  
  $\gamma$         & $1.65 \pm 0.1$      & $N(M) \sim L^\gamma$ \\
  $\alpha$         & $1.7 \pm 0.15$       & $M_\mathrm{cut} \sim L^\alpha, \xi^\alpha$ \\
  $\nu$            & $0.70 \pm 0.08$      & $\xi \sim \dot{\epsilon}^{-\nu}$ \\
 \hline 
\end{tabular}
	\caption{Reproduced estimates of critical exponents in 2D at $k_a = 2.5$ and their definitions from Ref. \cite{Clemmer2022} where $k_a = 0.0$. In 3D, $\tau = 1.7$ and $\phi = 0.7$. No significant dependence on material properties was identified in this article except in $\phi$.}
\label{table:exponents}
\end{table}

\subsection{Evolution with strain}
\label{sec:strain}

\begin{figure}
\begin{centering}
	\includegraphics[width=0.40\textwidth]{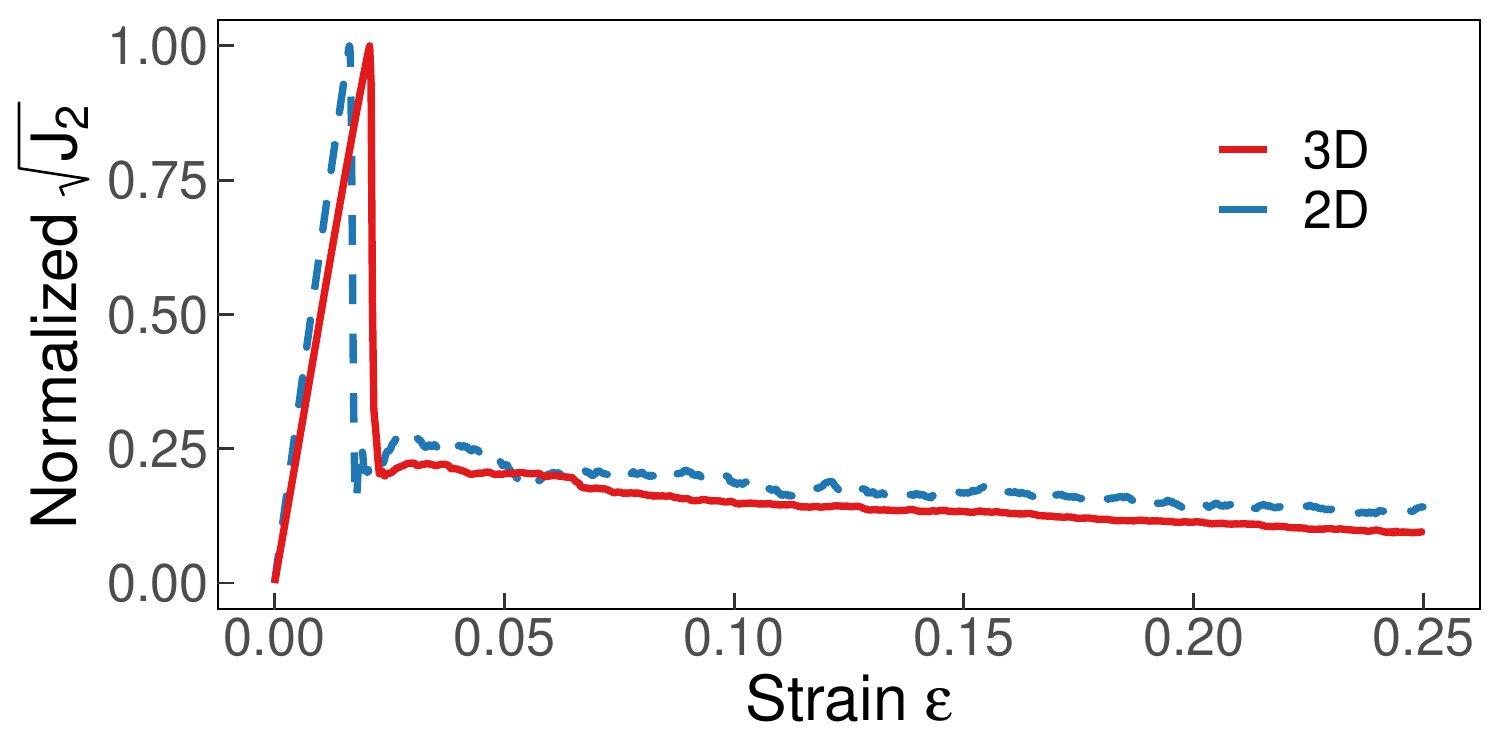}
	\caption{Shear stress as a function of strain in 2D (blue dashed line) and 3D (red solid line) normalized by the peak shear stress. In 2D, $L = 1600$, $k_a = 2.5$ ($\nu_\mathrm{PR} \sim 0.20$), and $\dot{\epsilon} = 10^{-5}$. In 3D, $L = 200$, $k_a = 3.0$ ($\nu_\mathrm{PR} \sim 0.17$), and $\dot{\epsilon} = 3\times10^{-5}$.} 
	\label{fig:stress_strain_qs}
\end{centering}
\end{figure}

\begin{figure*}
\begin{centering}
	\includegraphics[width=0.75\textwidth]{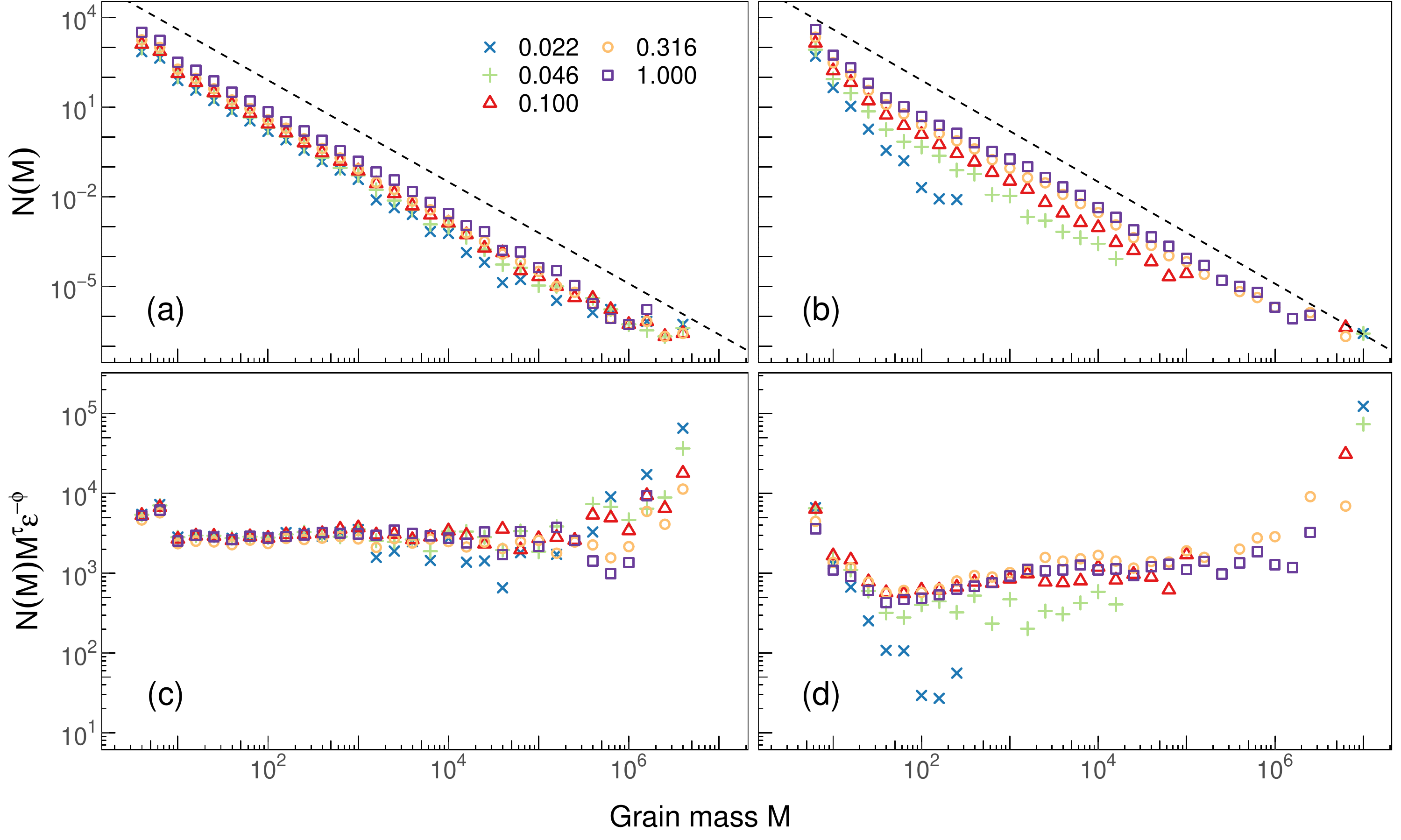}
	\caption{Distributions of the number of grains $N$ as a function of the grain mass $M$, $N(M)$, measured at the indicated strain (up to 1.0 or 100\%) in (a) and (c) 2D and (b) and (d) 3D using the same parameters in Fig. \ref{fig:stress_strain_qs}. In (c) and (d) distributions are scaled by $M^\tau \epsilon^{-\phi}$ with $\tau = 1.7$ in 2D and 3D and $\phi = 0.4$ in 2D and 0.7 in 3D. Dashed lines in (a) and (b) have slopes of $-\tau$. Dotted lines indicate estimates of $M_\mathrm{min} = 100$ (2D) and $1000$ (3D) and $M_\mathrm{max}$ is indicated by enlarged, outlined symbols.} 
	\label{fig:NM_strain}
\end{centering}
\end{figure*}

To begin, we characterize how systems fracture and fragment as they shear in the low-rate limit.
Results are presented in both 2D and 3D, although the discussion focuses on 2D systems which generally exhibit qualitatively similar behavior to 3D systems.
In Fig. \ref{fig:render_strain}, an example 2D system is rendered at various increments of strain to demonstrate the transition from a solid material to a highly polydisperse granular state as bonds are broken between particles.
Example stress-strain curves are plotted in Fig. \ref{fig:stress_strain_qs} where the shear stress is quantified as the square root of the second invariant of the deviatoric stress tensor, $\sqrt{J_2}$.

This system, as well as all others in this section, is strained at a relatively low rate of $10^{-5}$ in 2D and $3 \times 10^{-5}$ in 3D, which are quite close to the quasistatic limit and exhibit minimal finite-rate effects as further characterized in Sec. \ref{sec:fr}. 
In these figures and in the majority of demonstrative results in this text, we consider a single value of $k_a = 2.5$ or a Poisson's ratio $\nu_\mathrm{PR} \approx 0.20$ in 2D and $k_a = 3.0$ or $\nu_\mathrm{PR} \approx 0.17$ in 3D. 
This choice is used as a default to illustrate general behavior.
A nonzero value is chosen to contrast previous work which only used two-body interactions \cite{Clemmer2022}, while the specific value in 3D is chosen to reflect the elastic properties of boron carbide, an important ceramic for ballistic armor \cite{Ramesh2022}.
Note that the bulk modulus can be trivially adjusted to match that of boron carbide by defining the appropriate simulation units.

At zero strain, there are are no forces between particles and no stress in the system.
With increasing strain the stress grows linearly in Fig. \ref{fig:stress_strain_qs} before eventually dropping rapidly as the system brittlely fractures.
Shortly after fracture at a strain of 0.02 (or 2\%) there are a few large system-spanning cracks which cause failure, as seen in first panel of Fig. \ref{fig:render_strain}.
While these cracks may seem to have only broken the system into a few fragments or grains, there is actually significant structure and granular debris along the path of cracks, highlighted in the figure's inset.
During simulations, we track when each bond breaks to identify when new grains are produced.
A grain is defined as a disconnected subgraph in the bond network, i.e., an isolated set of bonded particles.
At regular intervals, the number of grains $N$ of a given mass $M$ is tallied to calculate the distribution of grain masses $N(M)$.
Grains smaller than $M = 3$ in 2D and $M = 4$ in 3D are not included in plots of $N(M)$.

Shortly after fracture, the system already contains grains with masses spread over six decades as seen in distributions $N(M)$ in Fig. \ref{fig:NM_strain}(a).
Furthermore, above some small threshold $M_\mathrm{min} \sim 100$, $N(M)$ resembles a power-law decay extending up to some maximum grain size cutoff $M_\mathrm{cut} \sim 10^5$.
Alternate behavior in small grains $M < M_\mathrm{min}$ is not surprising as fragmentation is ultimately limited by the size of a single particle, which affects statistics in this limit. 
At larger mass scales, the distribution curves slightly upward from the power law before reaching a maximum grain size of $M_\mathrm{max} \sim 3 \times 10^6$ or nearly half the mass of the entire system.
Grains of size $M_\mathrm{max}$ are exemplified by the large unbroken components in the first panel of Fig. \ref{fig:render_strain}.

As strain increases, the stress in Fig. \ref{fig:stress_strain_qs} then settles around a smaller value as the system undergoes granular flow.
During this flow regime, grains continue to fragment as demonstrated in the other panels of Fig. \ref{fig:render_strain} and the distribution shifts upward in Fig. \ref{fig:NM_strain}(a) as more grains of mass $M < M_\mathrm{max}$ are produced. 
Although it is not obvious in Fig. \ref{fig:NM_strain}(a), $M_\mathrm{cut}$ grows as the power law extends further (as further demonstrated below).
To provide the mass for this increase, $N(M_\mathrm{max})$ and $M_\mathrm{max}$ decrease as the largest fragments break up. 
This process continues until roughly one unit of strain where the power law extends up to the largest grains in the system and $M_\mathrm{cut} \sim M_\mathrm{max} \sim 10^6$.
In this limit, data is consistent with a power-law distribution $N(M) \sim M^{-\tau}$ with a value of $\tau = 1.7$, as previously measured in systems with $k_a = 0$ \cite{Clemmer2022}.
The power-law regime is highlighted in Fig. \ref{fig:NM_strain}(c) where distributions are normalized by $M^\tau$.
At larger strains, not shown, $M_\mathrm{cut}$ and $N(M)$ for $M > M_\mathrm{min}$ slowly decay as grains continue to break into smaller pieces.

In 3D systems, we see qualitatively similar behavior in Fig. \ref{fig:NM_strain}(b).
However, the distinction between $M_\mathrm{cut}$ and $M_\mathrm{max}$ is more pronounced and there is a clear gap that closes with increasing strain.
A power law is identifiable at $M > M_\mathrm{min} \sim 10^3$ starting at strains of $\epsilon \sim0.1$ and reaching a maximum span by $\epsilon = 1$.
The exponent is consistent with the previous estimate of $\tau = 1.7$ \cite{Clemmer2022}.
Simulations at other values of $k_a$ were not run in 3D due to computational costs.

Under a compaction loading geometry, experiments have found that comminution reaches a terminal state where the grains stabilize and stop breaking \cite{Altuhafi2011,Xiao2017}, as is often assumed in continuum models of breakage \cite{Einav2007, Einav2007b, Einav2007c, Cil2019, Bhattacharjee2021}. 
However, these results are hard to compare to our simulations due to the fundamental differences in loading.
A stronger comparison can be made to shear experiments by \citet{Marone1989} and \citet{Coop2004}, where grain breakage was found to persist to large strains before possibly reaching a steady-state limit at strains of $\sim 150\%$ and thousands of percent strain in the respective studies.
These approximate steady-state distributions were found to have fractal dimensions of approximately 2.5 or 2.6, which correspond to $\tau$ of $1.83$ or $1.87$.
This qualitative saturation in breakage at large strains and the estimates of exponents are reasonably close to the results from simulations studied here.

Compared to our previous results in larger 2D systems at lower rates at $k_a = 0$, we note two differences  \cite{Clemmer2022}.
In our prior work we observed a clear gap in the distribution for $M_\mathrm{cut} < M < M_\mathrm{max}$ for strains less than $\sim 1.0$ which is not present in the 2D data seen here (but is seen in 3D). 
However, a gap is also not seen in other 2D datasets run in this work at other values of $k_a$, including $k_a = 0$.
So this is not likely a physical change associated with the addition of three-body interactions but rather either due to the sampling statistics of these relatively rare grains or due to the slightly higher strain rate in these simulations.
More importantly, the other difference is the quantitative shift upward in distributions with increasing strain.
In Ref. \cite{Clemmer2022} we found that this growth was approximately proportional to strain to a power $\phi = 0.55 \pm 0.07$ such that the number of grains grew as $N(M) \sim \epsilon^\phi$ for $M_\mathrm{min} < M < M_\mathrm{cut}$.
Here the data is consistent with $\phi = 0.4 \pm 0.1$, as seen in Fig. \ref{fig:NM_strain}(c), where $N(M)$ is additionally  normalized by $\epsilon^\phi$.
This suggests that the increase in the number of grains with $\epsilon$ may depend on the linear elasticity of the system, $k_a$ or $\nu_\mathrm{PR}$.
In 3D, the vertical scaling of distributions with strain is consistent with $\phi = 0.7$, the value roughly estimated in Ref. \cite{Clemmer2022} at $k_a = 0.0$.
However, 3D data extends over an even narrower range of strains leading to more uncertainty about the presence of an actual power law and possibly requiring a larger range of $k_a$ to detect any dependence of $\phi$ on $k_a$.

\begin{figure}
\begin{centering}
	\includegraphics[width=0.40\textwidth]{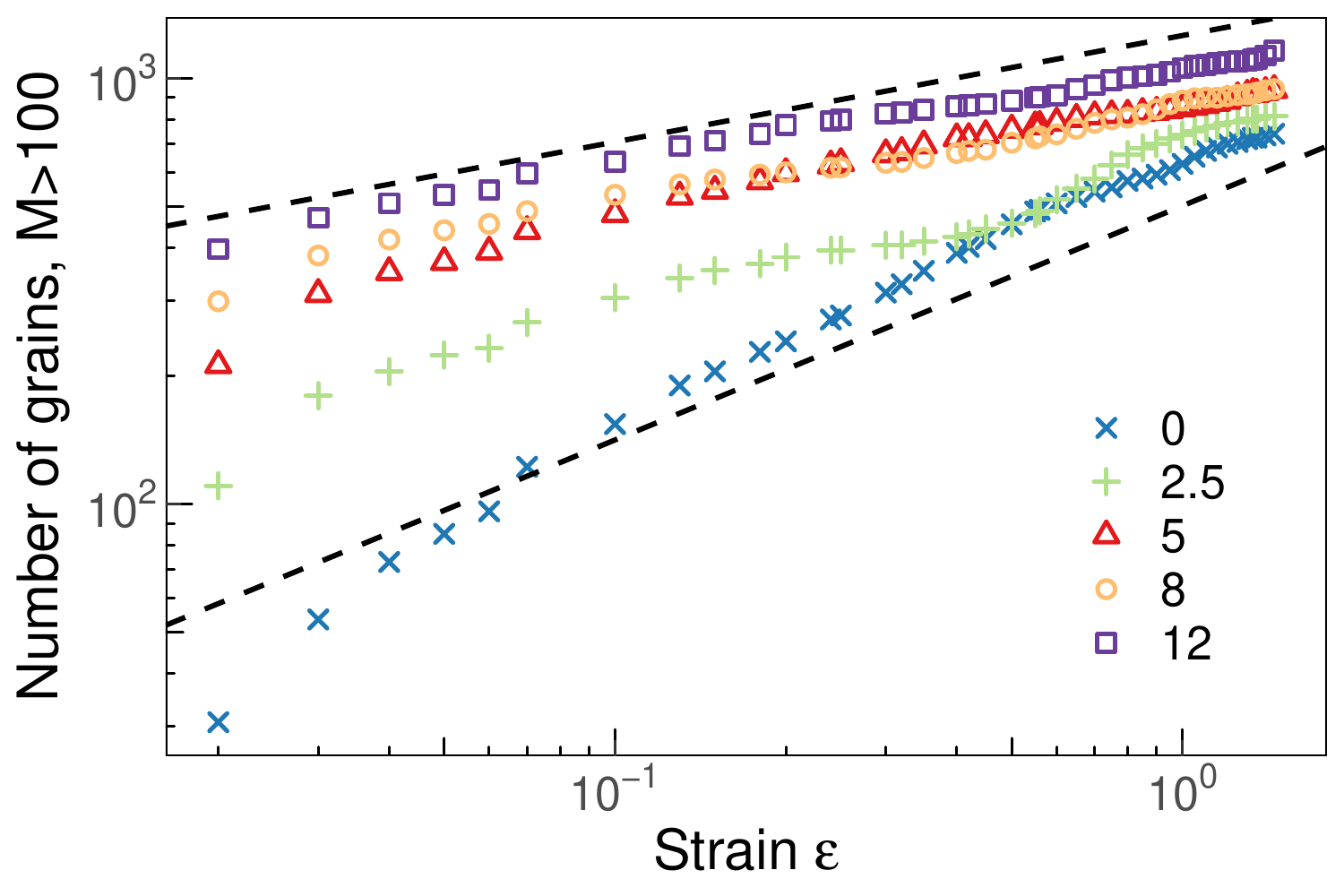}
	\caption{Number of grains with a mass $M > 100$ as a function of strain $\epsilon$ for the indicated values of $k_a$ in 2D. Values of $k_a = 0, 2.5, 5, 8,$ and $12$ correspond to $\nu_\mathrm{PR} \approx 0.33, 0.20, 0.10, 0.0,$ and $-0.11$, respectively. Dashed lines have slopes of 0.55 and 0.25.} 
	\label{fig:ngrain}
\end{centering}
\end{figure}

To further investigate the strain-dependence of the number of grains in 2D, we integrate distributions $N(M)$ in systems with $k_a$ ranging from zero to 12.0, or Poisson's ratios of $\nu_\mathrm{PR} = 1/3$ to -0.11 (well into the auxetic limit).
Only counting larger grains with masses $M > M_\mathrm{min}$, we find more grains are generated during the initial fracture of the system at large $k_a$ (small $\nu_\mathrm{PR}$).
However, new grains are created at a faster rate at small $k_a$ (large $\nu_\mathrm{PR}$), as seen in Fig. \ref{fig:ngrain}.
While we cannot determine whether this data reflects an actual power law due to the limited domain of $\epsilon$ considered and the fact that estimates of $\phi$ may depend on $M_\mathrm{min}$, one could feasibly measure an exponent $\phi$ from $\sim 0.55$ to $0.25$ with increasing $k_a$ (decreasing $\nu_\mathrm{PR}$), demonstrating a significant difference in the rate of breakage in systems with different Poisson's ratios. 

\begin{figure}
\begin{centering}
	\includegraphics[width=0.45\textwidth]{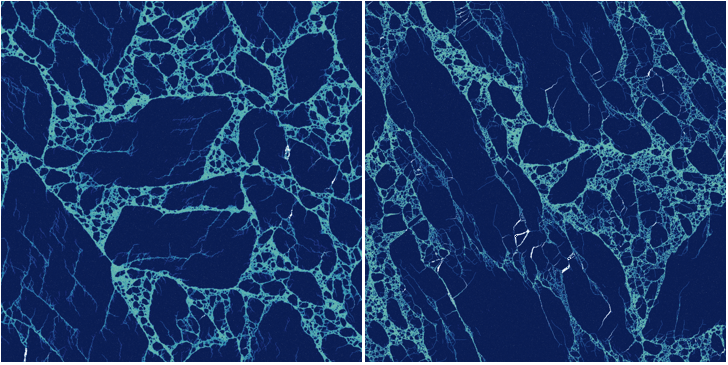}
	\caption{Rendered sections of a 2D system of size $L = 1600$ sheared to $\epsilon = 1.0$ at a faster rate of $\dot{\epsilon} = 3\times10^{-5}$ for values of $k_a = 0$ (left) and $12$ (right) or $\nu_\mathrm{PR} \approx 0.33$ and $-0.11$.} 
	\label{fig:render_k}
\end{centering}
\end{figure}

\begin{figure}
\begin{centering}
	\includegraphics[width=0.40\textwidth]{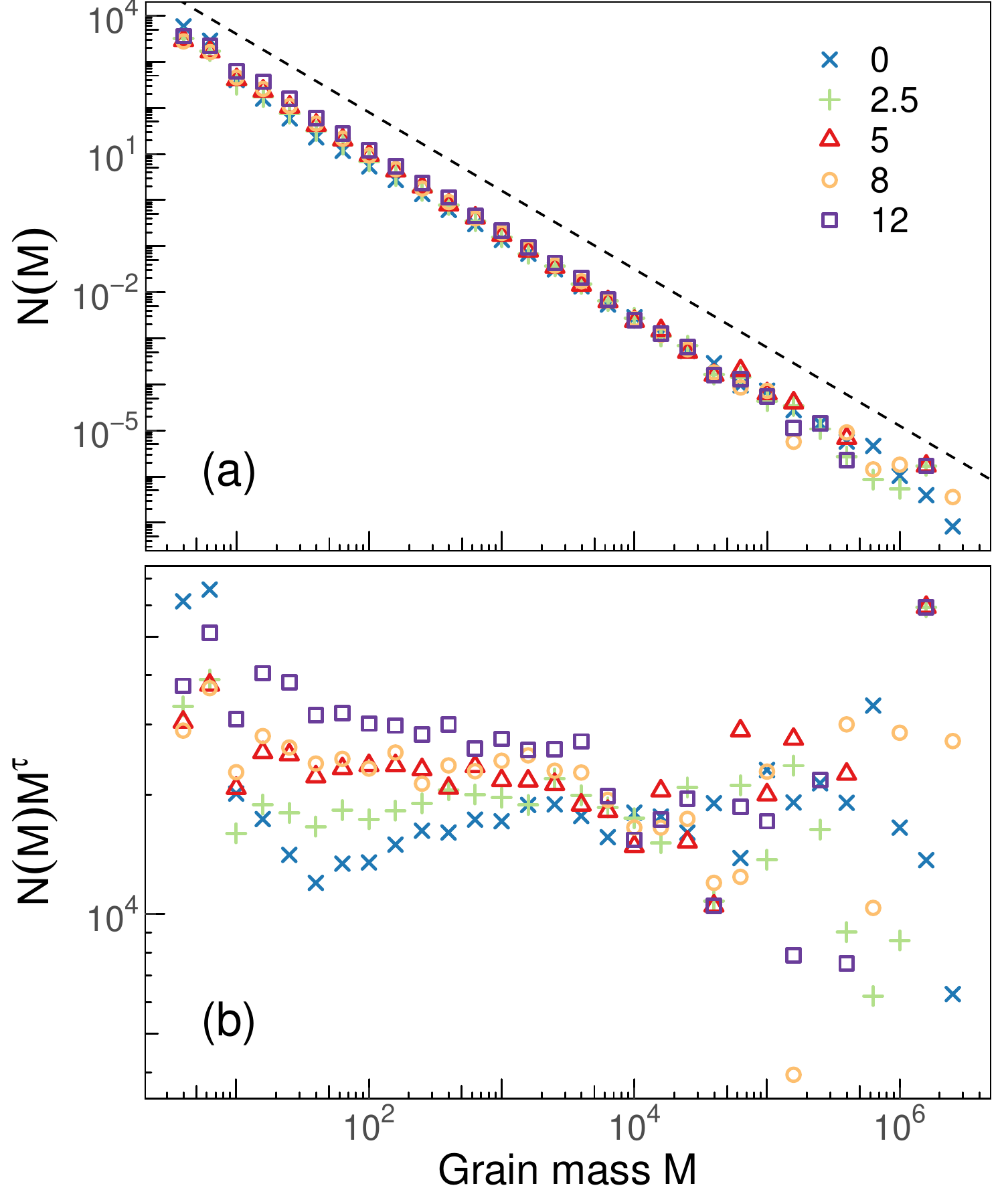}
	\caption{(a) Distributions $N(M)$ in 2D systems sheared to a strain of $\epsilon = 1.0$ at a rate of $\dot{\epsilon} = 10^{-5}$ for the indicated values of $k_a$. The dashed line has a slope of $\tau = 1.7$. (b) Same distributions as in (a) but scaled by $M^\tau$.} 
	\label{fig:NM_k}
\end{centering}
\end{figure}

Finally, we focus on the large-strain limit of $\epsilon = 1.0$ and explore the impact of varying $k_a$.
Poisson's ratio clearly affects the fragmentation process, as seen in Fig. \ref{fig:ngrain}, but does it affect the final critical state of the system?
Example fragmented systems at the extreme values of $k_a = 0$ and $12$ are seen in Fig. \ref{fig:render_k}.
Interestingly, there are observable differences in the shape of grains.
Grains appear more elongated with increasing $k_a$ (decreasing $\nu_\mathrm{PR}$). 
However, despite the changes both in the appearance of grains and in the evolution of $N(M)$ with strain, distributions of $N(M)$ at $\epsilon = 1.0$ are remarkably similar and have a minimal dependence on $k_a$ as seen in Fig. \ref{fig:NM_k}(a). 
Distributions all decay with $M$ with a power law close to $\tau = 1.7$.
One can identify a slight dependence on $k_a$ after dividing out the expected power law (Fig. \ref{fig:NM_k}[b]) as $N(M) M^\tau$ trends slightly downward with increasing $k_a$ (decreasing $\nu_\mathrm{PR}$).
However even in the extreme case of $k_a = 12$, one might only measure $\tau$ as high as 1.75 which is still within the estimated uncertainty in Ref. \cite{Clemmer2022}.
Additionally, this potential shift is most noticeable in grains of mass $M < 10^3$ and measurements of critical exponents should ideally focus on larger grains where a shift is not as clear.

A possible explanation for a slight shift in the power law of distributions could be that systems at different values of $k_a$ exhibit different lower mass scaling cutoffs $M_\mathrm{min}$ or different finite-size or finite-rate effects.
In the previous work in \cite{Clemmer2022}, the measured power-law exponent was found to vary with rate, as further discussed in Sec. \ref{sec:fr}.
To confirm the measured exponent does not change with decreasing rate, we collected data for $k_a = 12$ at an even slower rate of $3 \times 10^{-6}$ but found no significant difference.
Alternatively, it is possible that changes in the elasticity could shift the transition to the quasistatic limit in finite-size systems as further discussed in Sec. \ref{sec:frs}.
However, ultimately, any change in $\tau$ is still within uncertainty and cannot be determined to be significant.

\subsection{Finite-size effects}
\label{sec:size}

\begin{figure*}
\begin{centering}
	\includegraphics[width=0.9\textwidth]{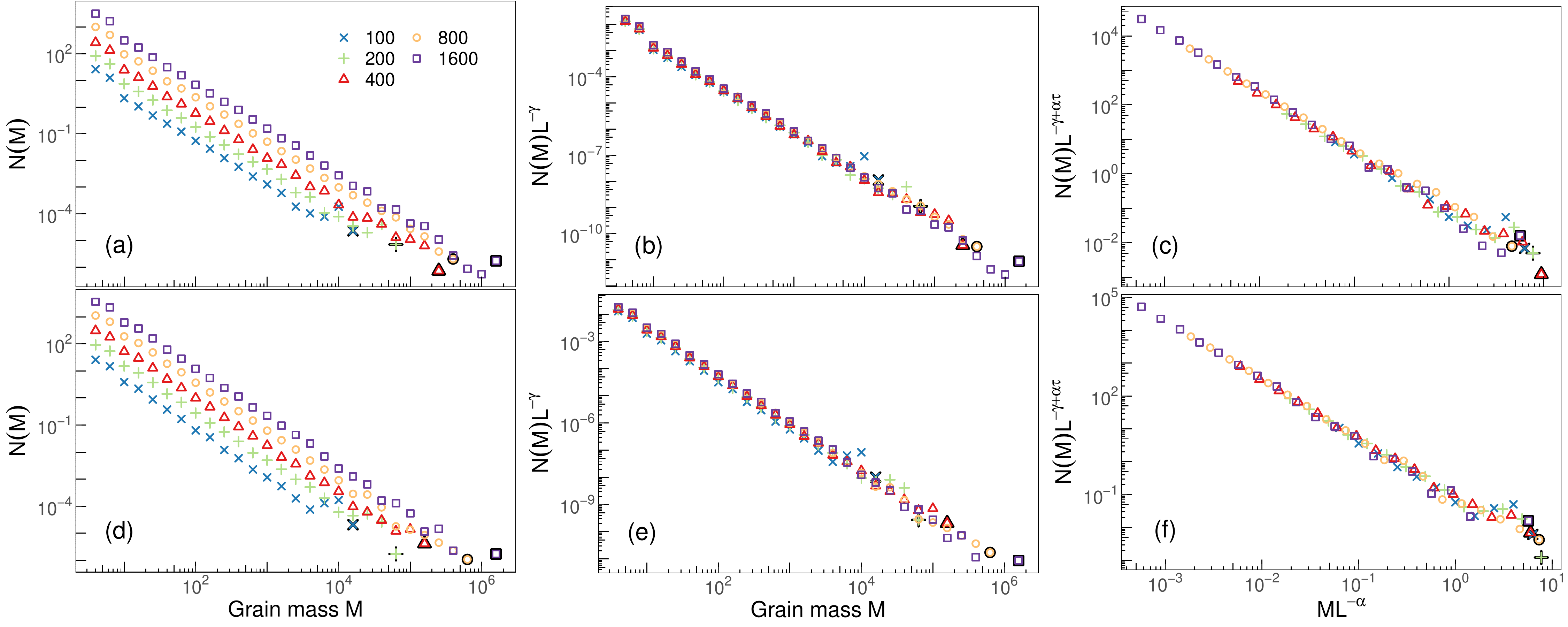}
	\caption{Distributions $N(M)$ in systems of the indicated size $L$ with (a)-(c) $k_a = 2.5$ and (d)-(f) $k_a = 12$ sheared to a strain of $\epsilon = 1.0$ at a rate of $\dot{\epsilon} = 10^{-5}$ for $L = 1600$ and $\dot{\epsilon} = 3 \times 10^{-5}$ for all other $L$: (a) and (d) raw distributions, (b) and (e) $N(M)$ scaled by $L^\gamma$, and (c) and (f) a fully scaled $N(M)$ using Eq. \eqref{eq:nqs} plotted using values of $\tau = 1.7$, $\gamma = 1.65$, and $\alpha = 1.7$. For each data set, the largest mass at each $L$ is enlarged and outlined to improve visibility.} 
	\label{fig:NM_k_size}
\end{centering}
\end{figure*}

In smaller systems, one naturally expects fewer and smaller grains.
This is seen for 2D systems in Fig. \ref{fig:NM_k_size} at both $k_a = 2.5$ or $\nu_\mathrm{PR} \approx 0.20$ in Fig. \ref{fig:NM_k_size}(a) and $k_a = 12$ or $\nu_\mathrm{PR}  \approx -0.11$ in Fig. \ref{fig:NM_k_size}(d).
Systems are sheared to a strain of 1.0 at a rate of $\dot{\epsilon} = 10^{-5}$ for the largest system size of $L = 1600$ and $\dot{\epsilon} = 3 \times 10^{-5}$ for all other sizes.
At both values of $k_a$, $N(M)$ shifts downward with decreasing linear system size $L$ while the upper cutoff of the power law $M_\mathrm{cut}$ similarly drops.

A reasonable assumption might be that the number of grains should scale with the total mass in the system such that $N(M) \sim L^d$, where $d$ is the spatial dimension.
However, while the vertical shift in distributions is reasonably described by a power law $N(M) \sim L^\gamma$, the exponent $\gamma$ is distinctly less than $d$ for both values of $k_a$ as measured by collapsing distributions in Figs. \ref{fig:NM_k_size}(b) and \ref{fig:NM_k_size}(e) using a value of $\gamma = 1.65 \pm 0.1$ from previous work \cite{Clemmer2022}.
Notably, this implies the number of grains grows subextensively with the size of the system.
After scaling, one can still identify a slight splay in distributions across $L$ such that data may be better fit by $\gamma = 1.75$ for all $k_a$ studied in this work except for $k_a = 12.0$, which may be better fit by $\gamma = 1.80$.
However, like estimates of $\tau$, this determination depends on which domain is considered, ideally focusing on the largest grains for which statistics are weakest.
Therefore, we cannot conclude this effect is significant, again noting that these results rely on overall smaller systems at higher rates with fewer statistics than previous work.

To capture the dependence of $N(M)$ on $L$, we construct a finite-size scaling theory for $N(M)$.
This process is based on similar derivations of critical scaling theories for the magnitude of avalanches in the depinning and yielding transitions \cite{Ji1992, Perkovic1999, Salerno2012, Salerno2013, Lin2014, Clemmer2019}.
We assume that $M_\mathrm{cut} \sim L^\alpha$, where $\alpha$ is a critical exponent, and that $N(M)$ only depends on the ratio $M/M_\mathrm{cut}$.
This leads to the ansatz
\begin{equation}
N_\mathrm{QS}(M,L) = L^{\gamma-\alpha \tau} f(M/L^\alpha) \ \ ,
\label{eq:nqs}
\end{equation}
where $f(x)$ is a scaling function.
To ensure there are no grains larger than $M_\mathrm{cut}$, $f(x)$ must go to zero for $M \gg L^\alpha$ or $x \gg 1$.
In the opposite limit of $M \ll L^\alpha$, $f(x)$ must scale as $x^{-\tau}$ for $x \ll 1$ such that $N_\mathrm{QS} \sim L^\gamma M^{-\tau}$.
Using this scaling relation, we find distributions are reasonably collapsed using a value of $\alpha = 1.7$ in Figs. \ref{fig:NM_k_size}(c) and \ref{fig:NM_k_size}(f).
Notably, this implies the size of the largest grain also grows sub-extensively with the size of the system.
There is some splay in the scaled data near the cutoff $M_\mathrm{cut}$ which might be due to noncritical behavior in smaller systems or slight finite-rate effects in large systems, as discussed in the following sections. 

As it is difficult to rigorously identify the location of the cutoff $M_\mathrm{cut}$ in Fig. \ref{fig:NM_k_size} due to limited statistics of large grains, we alternatively consider the moments of the distribution to improve estimates of $\alpha$ and derive a scaling relation between exponents.
Calculating the $n^\mathrm{th}$ moment, we find
\begin{align}
\langle M^n \rangle_\mathrm{QS} 
     & = \int M^n N_\mathrm{QS}(M,L) dM \\
     & = \int L^{\gamma-\alpha \tau} M^n f(M/L^\alpha) dM \ \ .
\end{align}
Note that this definition is not normalized by the total number of grains.
Substituting variables for $x = M/L^\alpha$ yields
\begin{equation}
\langle M^n \rangle_\mathrm{QS}  = L^{\gamma+\alpha(n + 1 - \tau)} \int x^n f(x) dx
\label{eq:moment_qs}   
\end{equation}
where the integral is dominated by the upper limit and converges for $n > \tau - 1 \sim 0.7$.
Since this expression is not normalized, the lowest moment $n = 1$ simply equals the total mass in the system and scales as $L^d$ in $d$ dimensions.
This implies a scaling relation
\begin{equation}
d = \gamma + \alpha (2 - \tau) \ \ .
\label{eq:scaling_relation}   
\end{equation}
A similar scaling relation exists for exponents describing the distribution of avalanche magnitudes in sheared disordered systems in the yielding transition \cite{Salerno2012, Salerno2013, Clemmer2021a}.

\begin{figure}
\begin{centering}
	\includegraphics[width=0.40\textwidth]{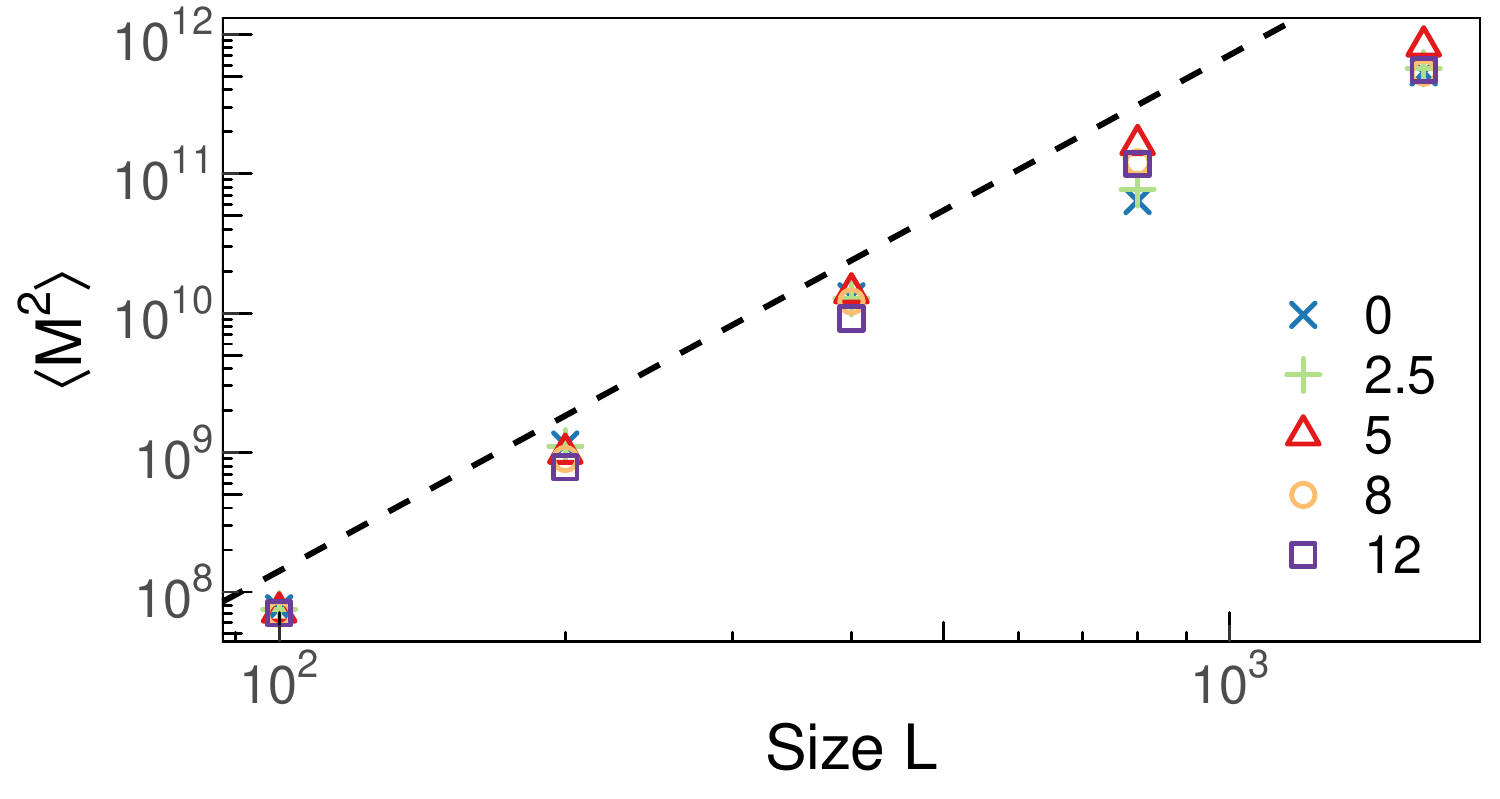}
	\caption{Mean-square grain size as a function of 2D system size $L$ for the indicated values of $k_a$. Data is collected at a strain of $\epsilon = 1.0$ at a rate of $\dot{\epsilon} = 10^{-5}$ for $L = 1600$ and $\dot{\epsilon} = 3 \times 10^{-5}$ for all other sizes. The dashed line represents a power law with exponent $2+ \alpha$ with $\alpha = 1.7$.} 
	\label{fig:msq}
\end{centering}
\end{figure}

For the second moment we find
\begin{equation} 
\langle M^2 \rangle \sim L^{\gamma + \alpha (3 -\tau)} \sim L^{d + \alpha}
\label{eq:moment2_qs}
\end{equation}
using the scaling relation in Eq. \eqref{eq:scaling_relation}.
This expression is particularly useful as it isolates the exponent $\alpha$.
Calculating the second moment from the same data used to produce Fig. \ref{fig:NM_k_size}, we find $\langle M^2 \rangle$ grows with system size $L$ at all values of $k_a$ (Fig. \ref{fig:msq}).
This growth is consistent with a power-law exponent $\alpha = 1.7$ measured in Ref. \cite{Clemmer2022}.
The value of $\langle M^2 \rangle$ is overpredicted in the largest system size, which could imply a smaller value of $\alpha$ at all $k_a$, however, the possible deviation in $\alpha$ is not larger than the $\pm 0.15$ range of uncertainty estimated in the previous work (Table \ref{table:exponents}).
Furthermore, as noted before, the largest system sizes may have slight finite-rate effects leading to a smaller values of $\langle M^2 \rangle$.
In the following two sections, we incorporate the scaling of $\langle M^2 \rangle$ with rate to account for this possibility.

\subsection{Finite-rate effects}
\label{sec:fr}

\begin{figure}
\begin{centering}
	\includegraphics[width=0.40\textwidth]{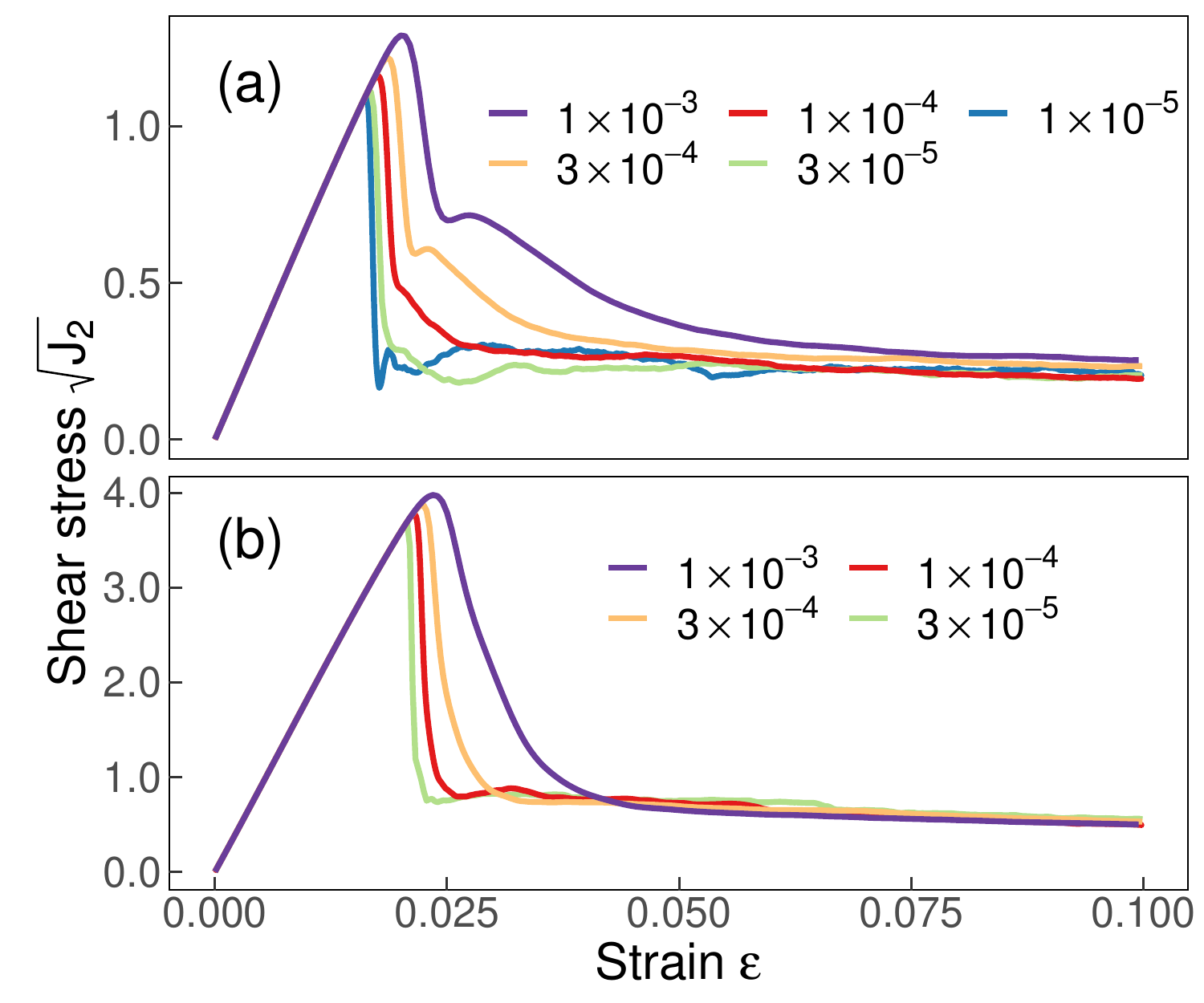}
	\caption{Stress ratio as a function of strain at the indicated strain rates $\dot{\epsilon}$ in (a) 2D at $L = 1600$ and $k_a = 2.5$ and (b) 3D at $L = 200$ and $k_a = 3.0$.} 
	\label{fig:stress_strain}
\end{centering}
\end{figure}

Up to now, results have focused on the low-strain-rate limit. 
With increasing strain rate, systems yield at larger strains and fracture occurs over a wider range of strain, as seen in Fig. \ref{fig:stress_strain}. 
At larger strains, the shear stress still stabilizes as the system reaches a quasi-steady state of granular flow and comminution; however, systems stabilize around larger stress at higher rates.
This is clearly visible in 2D stress-strain curves but also occurs in 3D, although the magnitude of the effect is smaller and is only easily visible at larger strains.
As in the quasistatic limit, the flow stress gradually decays with increasing strain.

\begin{figure*}
\begin{centering}
	\includegraphics[width=0.98\textwidth]{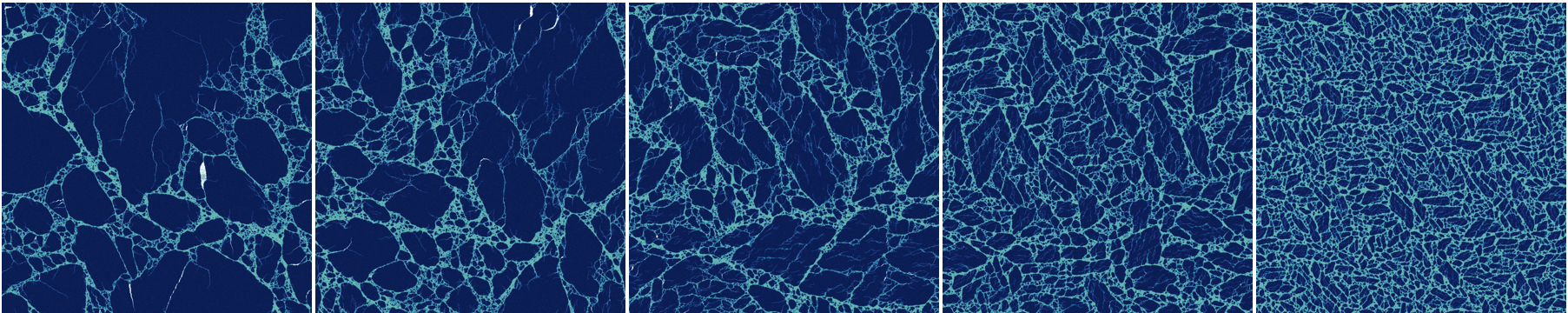}
	\caption{Rendered sections of a fragmented 2D system of size $L = 1600$ with $k_a = 2.5$ sheared to a strain of $\epsilon = 1.0$ at rates of $\dot{\epsilon} = 10^{-5}, 3\times10^{-5}, 10^{-4}, 3 \times 10^{-4}$, and $10^{-3}$ going from left to right.} 
	\label{fig:render_rate}
\end{centering}
\end{figure*}

\begin{figure*}
\begin{centering}
	\includegraphics[width=0.75\textwidth]{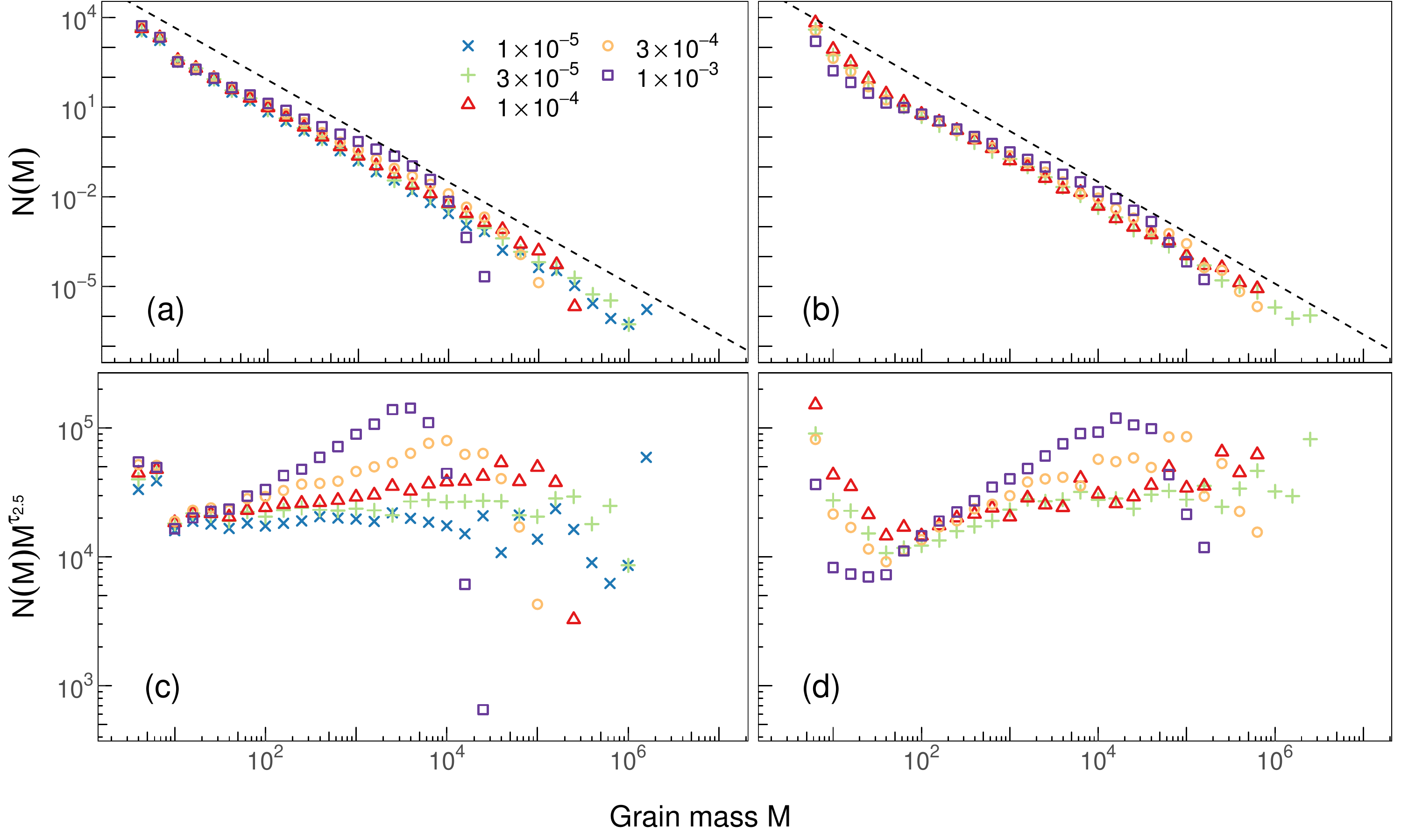}
	\caption{Distributions $N(M)$ in systems of size $L=1600$ sheared to a strain of 1.0 at the indicated rates in (a) and (c) 2D with $k_a = 2.5$ and (b) and (d) in 3D with $k_a = 3.0$. In (c) and (d) distributions are scaled by $M^\tau$ with $\tau = 1.7$. Dashed lines in (a) and (b) have slopes of $-\tau$.} 
	\label{fig:NM_2d_3d_rate}
\end{centering}
\end{figure*}

These changes in the mechanical response arise from the underlying changes in the microscopic granular structure.
In Fig. \ref{fig:render_rate} a dramatic change in the characteristic size of grains with increasing strain rate is seen in snapshots of 2D systems sheared to one unit of strain.
As one would expect, at faster strain rates the system nucleates and grows more cracks \cite{Cereceda2017}, causing the system to break into smaller fragments \cite{Grady1985, Lankford1996, Astrom2000, Zhou2006, Wittel2008, Carmona2008, Levy2010}.
Here we aim to quantify how the characteristic size of grains decreases over this broad span of strain rates.

To quantify this behavior, we again turn to distributions of grain masses $N(M)$ as seen in both 2D and 3D in Fig. \ref{fig:NM_2d_3d_rate}.
As the strain rate increases, the upper limit of $N(M)$ shrinks reflecting the qualitative reduction in the largest grain size seen in Fig. \ref{fig:render_rate}.
While $N(M)$ still appears to decay as a power of $M$ at high rates, interestingly $N(M)$ becomes less steep with increasing rate.
This trend becomes more apparent after normalizing by the quasistatic power law (Fig. \ref{fig:NM_2d_3d_rate}[c-d]) as $N(M) M^\tau$ clearly rises above the predicted quasistatic power law at high rates.
This data could suggest either that the exponent $\tau$ systematically depends on the strain rate, which would be quite unusual, or that the actual power law only emerges for rates smaller than $\sim 10^{-4}$ such that higher rates exhibit some other non-critical behavior.
As the difference between exponents of two consecutive curves decreases with progressively decreasing rates, the slope appears to be converging, suggesting the data is more consistent with the latter option.
However, resolving this distinction would again require running simulations of even larger systems at slower rates.

\begin{figure*}
\begin{centering}
	\includegraphics[width=0.75\textwidth]{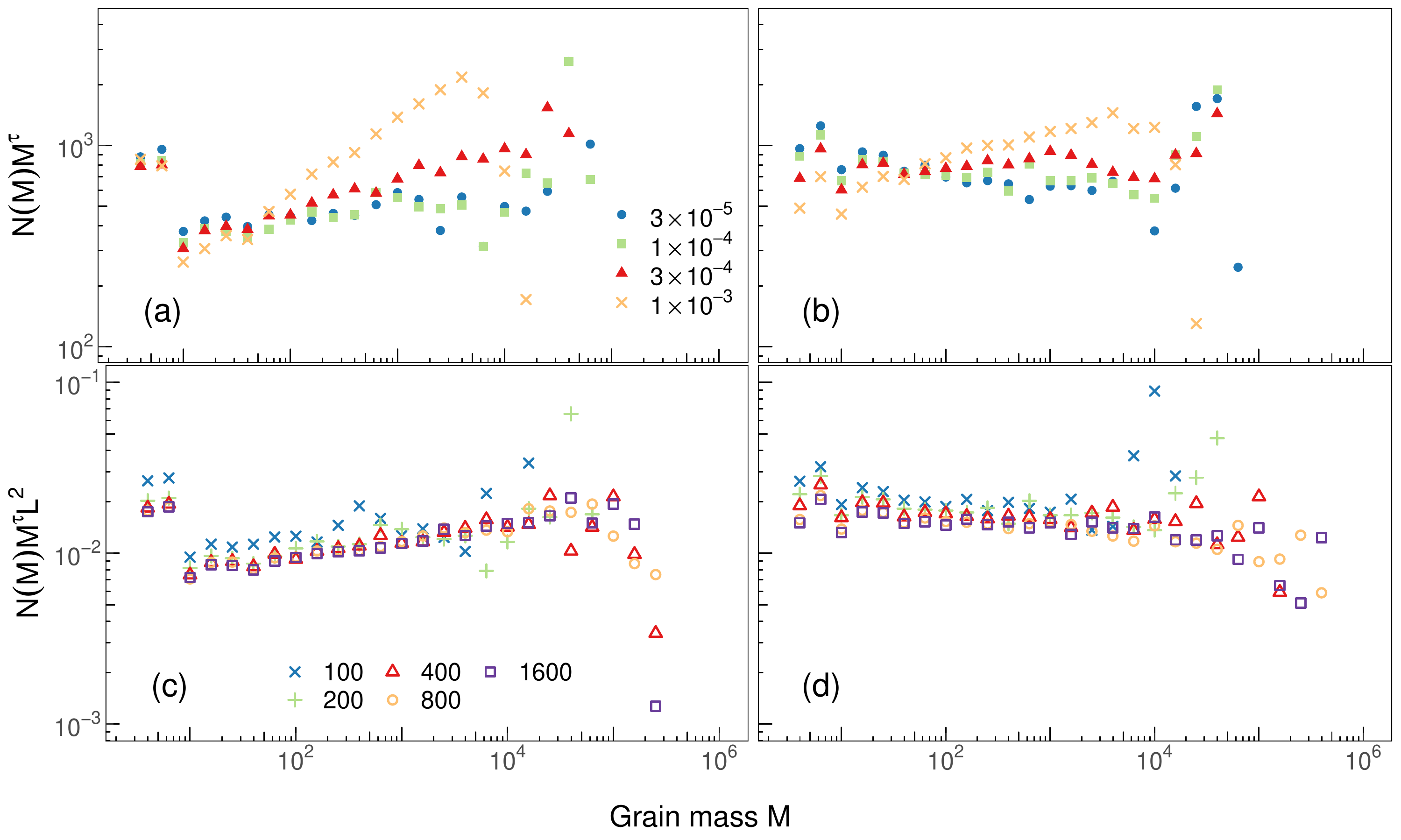}
	\caption{(a) and (b) Mass distributions scaled by $M^{\tau}$ in 2D systems of size $L = 200$ sheared at the indicated rates. (c) and (d) Distributions scaled by $M^{\tau} L^2$ for the indicated system sizes sheared at a fixed rate of $10^{-4}$. Normalization uses a value of $\tau = 1.7$. All systems are sheared to $\epsilon = 1.0$ and correspond to (a) and (c) $k_a = 2.5$ and (b) and (d) $k_a = 12$. } 
	\label{fig:NM_size_rate}
\end{centering}
\end{figure*}

Regardless of the origin of this effect, it is interesting to note that one could conceivably measure a wide range of $\tau$ spanning from 1.7 to around 1.3 in 2D and 3D depending on rate.
This could reflect the range of exponents measured in different experimental systems \cite{Turcotte1986}.
While the quasistatic power law of $N(M)$ is found to be quite robust to changes in material properties, evidencing universal behavior within the regime of isotropic brittle materials, one could potentially still measure different exponents at high rates. 
Although this paper does not include a thorough dissection of the evolution of $N(M)$ with strain at high rates, we note that $N(M)$ evolves very differently with strain compared to the low-rate limit discussed in Sec. \ref{sec:strain}.
Specifically, at high rates the distribution is initially steeper at small strains but becomes less steep with increasing strain up to strains of $\sim 1.0$.
In a system sheared at a high rate, one could therefore potentially fit different exponents depending on the strain.
This behavior was discussed briefly in Ref. \cite{Clemmer2022} for $k_a = 0.0$, and similar trends were identified in systems with $k_a \ne 0.0$, although further simulations and alternate analytic techniques like population balance modeling are required to thoroughly explore this behavior.

With increasing strain rate, energy is added more rapidly to the system.
To confirm that energy dissipates quickly enough and that results reflect the overdamped limit, we ran simulations at the fastest rate of $\dot{\epsilon} = 10^{-3}$ with different damping strengths $\gamma$ between 5.0 and 500.0 where the standard value of $\gamma$ is $50.0$. 
With changing $\gamma$, there is a systematic change in the number of small grains $M < M_\mathrm{min}$, varying up to $\pm 50\%$ relative to results at $\gamma = 50.0$.
This splay converges with increasing $\gamma$. 
Additionally, the minimum cutoff of the power law $M_\mathrm{min}$ increases to $\sim 1000$ at the smallest value of $\gamma = 5.0$. 
However, there is no significant change in the number of grains for $M > M_\mathrm{min}$. 
A similar power law can be fit to $N(M)$ at all $\gamma$ and the upper cutoff $M_
\mathrm{cut}$ is consistent, implying our results are representative of the overdamped limit.
At the second fastest rate of $\dot{\epsilon} = 3 \times 10^{-4}$, there is minimal dependence on $\gamma$ even in the small-mass limit for $\gamma \in [5.0, 500.0]$. 
However, it is likely that even smaller values of $\gamma$ could change results.

\subsection{Finite-size and finite-rate effects}
\label{sec:frs}

Having separately considered finite-size and finite-rate effects in the preceding two sections., we now consider the combination to determine the crossover between the finite-rate and quasistatic limits and its scaling with system size.
As before, we continue to focus on the large-strain limit of $\epsilon = 1.0$ in 2D systems.
In Fig. \ref{fig:NM_size_rate}, normalized distributions from a relatively small system of size $L = 200$ are plotted at different strain rates at $k_a = 2.5$ ($\nu_\mathrm{PR} \sim 0.20$, Fig. \ref{fig:NM_size_rate}[a]) and $12$ ($\nu_\mathrm{PR} \sim -0.11$, Fig. \ref{fig:NM_size_rate}[b]). 
Similar to Fig. \ref{fig:NM_2d_3d_rate}(c), distributions have a shallower power law and are truncated at smaller masses at the fastest rates. 
However, as the rate decreases to $10^{-4}$ and below, distributions cease evolving, suggesting this smaller system size has already reached the quasistatic limit, at a higher rate than simulations of size $L=1600$. 

If we alternatively focus on a higher strain rate of $\dot{\epsilon} = 10^{-4}$ and vary the system size, we see different behavior between larger systems, $L > 400$, and smaller systems, $L < 400$. 
Large systems exhibit a trivial dependence on $L$ as distributions simply scale extensively, $N(M) \sim L^2$, and have similar cutoffs as seen in Figs. \ref{fig:NM_size_rate}(c) and \ref{fig:NM_size_rate}(d).
In contrast, there is some vertical splay in distributions for small systems and the cutoff grows with increasing $L$.
This phenomenon is consistent with the above finding that smaller systems transition to the quasistatic limit at higher strain rates and is naturally anticipated.
For instance, one might expect the transition to quasistatic behavior could be governed by the timescale it takes for a crack to propagate across the system such that the crossover occurs at a rate of $\dot{\epsilon} \sim L^{-1}$ for a constant crack propagation speed.
However, as identified below, this criterion is not sufficient.

To capture the transition and quantify how it scales with $L$, we hypothesize that there is an additional length scale $\xi$ that diverges with decreasing strain rate $\dot{\epsilon}$ with an exponent $\nu$:
\begin{equation}
    \xi \sim \dot{\epsilon}^{-\nu} \ \ .
    \label{eq:xi}
\end{equation} 
Based on the findings from Fig. \ref{fig:NM_size_rate}, we assume fragmentation depends only on the smallest of the two length scales $L$ and $\xi$. 
At low rates where $\xi > L$, the system is in the quasistatic limit such that $N(M)$ is described by Eq. \eqref{eq:nqs} and the size of the largest grain $M_\mathrm{cut}$ is set by $L^\alpha$.
At high rates where $\xi < L$, the system is in the finite-rate limit and $\xi$ governs behavior such that $M_\mathrm{cut} \sim \xi^\alpha$.
This idea echoes results from many dynamic critical phenomena such as avalanches in depinning and yielding \cite{Narayan1993, Sethna2001, Chauve2001, White2003, Lin2014, Clemmer2021a, Clemmer2021b}.

Next we propose a scaling description for $N(M)$ in the finite-rate limit.
From Figs. \ref{fig:NM_size_rate}(c) and \ref{fig:NM_size_rate}(d) we assume $N(M)$ primarily depends on $\xi$ and only has a trivial dependence on $L$ in this limit, namely, $N(M)$ grows extensively as $L^d$ in $d$ dimensions.
A system of size $L$ can then be divided up into $(L/\xi)^d$ independent regions, each of size $\xi^d$.
Within each region, the distribution of grains resembles that of a quasistatic system of size $L = \xi$, or $N_\mathrm{QS}(M, L = \xi)$ from  Eq. \eqref{eq:nqs}.
Combining the contributions from all of these regions yields a finite-rate ansatz for $N(M)$ of the entire system,
\begin{equation}
N_\mathrm{FR}(M,\dot{\epsilon},L)  \sim L^d \dot{\epsilon}^{\nu(d-\gamma+\alpha \tau)} h (M \dot{\epsilon}^{ \nu \alpha }) \ \ ,
\end{equation}
where $h(x)$ is another scaling function.
If $x \gg 1$, $h(x)$ goes to zero while if $x \ll 1$ then $h(x) \sim x^{-\tau}$ such that $N_\mathrm{FR}(M,\dot{\epsilon}) \sim L^d \dot{\epsilon}^{\nu(d-\gamma)} M^{-\tau}$. 
Using the scaling relation for $\gamma$ in Eq. \eqref{eq:scaling_relation}, this ansatz can be reexpressed as
\begin{equation}
N_\mathrm{FR}(M,\dot{\epsilon},L)  \sim L^d \dot{\epsilon}^{2 \nu \alpha} h (M \dot{\epsilon}^{ \nu \alpha }) \ \ .
\label{eq:nfr}
\end{equation}
Note that this derivation does not account for the observation that distributions become more shallow at high strain rates or that $\tau$ may possibly depend on $\dot{\epsilon}$.
As the magnitude of this deviation appears to decrease at lower strain rates in larger systems, we postulate that it will not be significant in the large-system low-rate limit and therefore neglect it, although further validation is needed.

To test this theory, we again turn towards the moments of the distribution. 
As in Sec. \ref{sec:size}, one can derive an expression for the moments of $N(M)$ in the finite-rate limit as
\begin{align}
\langle M^n \rangle_\mathrm{FR} 
     & = \int M^n N_\mathrm{FR}(M,\dot{\epsilon},L) dM \\
     & = \int L^d \dot{\epsilon}^{2 \nu \alpha} M^n h (M \dot{\epsilon}^{ \nu \alpha }) dM \ \ .
\end{align}
Again, substituting variables for $x = M \dot{\epsilon}^{ \nu \alpha }$ yields
\begin{equation}
\langle M^n \rangle_\mathrm{FR}  = L^d \dot{\epsilon}^{\nu \alpha (1 - n)} \int x^n h(x) dx
\label{eq:moment_fr}   
\end{equation}
where the integral is similarly dominated by the upper limit and converges for $n > \tau - 1 \sim 0.7$.
For the second moment, one finds
\begin{equation}
\langle M^2 \rangle_\mathrm{FR}  \sim L^d \dot{\epsilon}^{-\nu \alpha} \ \ .
\label{eq:moment2_fr}   
\end{equation}
Combining this with the expression for $\langle M^2 \rangle_\mathrm{QS}$ in Eq. \eqref{eq:moment2_qs}, we can now construct a scaling ansatz for $\langle M^2 \rangle$ across system sizes and rates.
We assume that $\langle M^2 \rangle$ only depends on the dimensionless ratio of $L/\xi$ or $L \dot{\epsilon}^\nu$ such that
\begin{equation}
\langle M^2 \rangle \sim L^{d + \alpha} g(\dot{\epsilon} L^{1/\nu}) \ \ ,
\label{eq:ansatz_m2}
\end{equation}
where $g(x)$ is yet another scaling function.
In the quasistatic limit where $x \ll 1$, $g(x)$ goes to a constant to recover the scaling in Eq. \eqref{eq:moment2_qs}.
In the finite-rate limit where $x \gg 1$, $g(x) \sim x^{-\nu \alpha}$ to recover the scaling in Eq. \eqref{eq:moment2_fr}. 

\begin{figure}
\begin{centering}
	\includegraphics[width=0.40\textwidth]{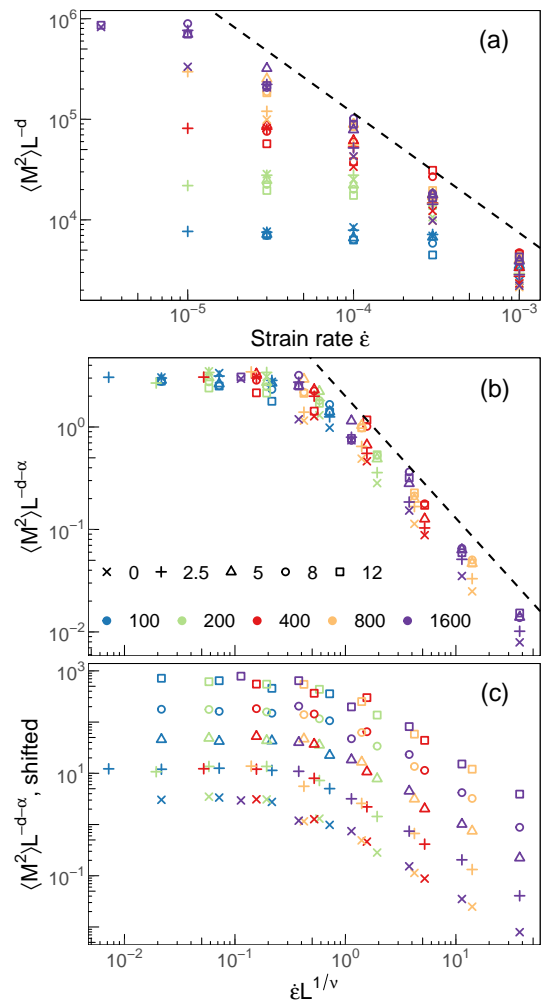}
	\caption{(a) Second moment of $N(M)$ normalized by $L^d$ with $d = 2$ as a function of strain rate for different 2D system sizes $L$ (color) and values of $k_a$ (shape) as indicated by the legend in (b). The dashed line represents a power law with an exponent $-\nu \alpha$. (b) The same data in (a) is collapsed using the scaling relation in Eq. \eqref{eq:ansatz_m2}. (c) The scaled data in (b) is plotted after shifting data vertically across values of $k_a$ for visibility. All panels use values of $\alpha = 1.7$ and $\nu = 0.7$.} 
	\label{fig:msq_collapse}
\end{centering}
\end{figure}

Calculating $\langle M^2 \rangle$ across different system sizes and rates, we see behavior consistent with the theory.
In Fig. \ref{fig:msq_collapse}(a) in the high-rate limit, $\langle M^2 \rangle$ grows as a power of decreasing rate at all $L$ independent of the elastic properties or $k_a$. 
This growth continues until $\dot{\epsilon} \sim 3 \times 10^{-4}$ where $\langle M^2 \rangle$ plateaus for the smallest systems $L = 100$ and no longer depends on rate.
A similar crossover is then seen in progressively larger systems at lower and lower rates as systems transition to the quasistatic limit.
Comparing different values of $k_a$, a similar trend is seen although the location of the crossover and the height of the plateau may shift.

\begin{figure*}
\begin{centering}
	\includegraphics[width=0.98\textwidth]{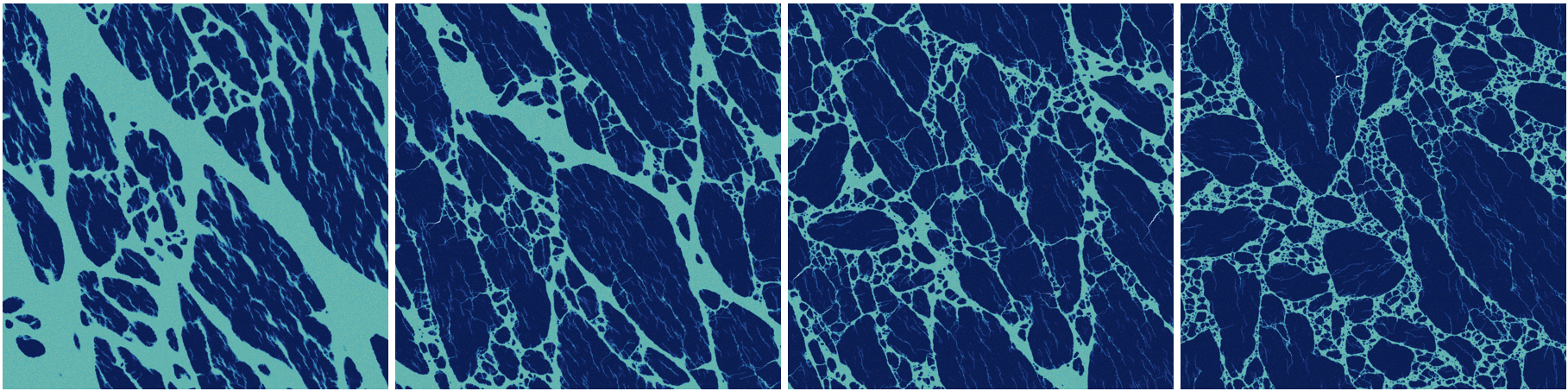}
	\caption{Rendered sections of a 2D system of size $L = 1600$ sheared to a strain of 1.0 at a faster rate of $3\times10^{-5}$ with values of $k_a = 2.5$, $\theta_c = 10^\circ$, and $\lambda_c = 1.01, 1.015, 1.02,$ and $1.03$ in panels going from left to right.} 
	\label{fig:render_lc}
\end{centering}
\end{figure*}

Scaling data by the system size according to Eq. \eqref{eq:ansatz_m2}, data is collapsed using values of $\alpha = 1.7$ and $\nu = 0.7$ from Ref. \cite{Clemmer2022} as shown in Fig. \ref{fig:msq_collapse}(b).
There is some observable splay which may simply be due to different constant prefactors (such as a vertical or horizontal shift) across values of $k_a$.
Shifting datasets apart in Fig. \ref{fig:msq_collapse}(c) reveals that the quality of the individual collapses is reasonable across values of $k_a$.
Focusing on a single dataset, one might estimate a somewhat different exponent depending on the value of $k_a$; however, these deviations are not greater than uncertainty in exponents.

\subsection{Impact of fracture toughness}
\label{sec:kc}

In the previous sections, we considered how changes in the elastic properties of the material affect fragmentation.
This is accomplished by varying $k_a$ which adjusts Poisson's ratio $\nu_\mathrm{PR}$.
In this section, 2D systems with a fixed value of $k_a = 2.5$ or $\nu_\mathrm{PR} \sim 0.20$ are used to alternatively test how fracture toughness impacts fragmentation.
We consider variations in $K_\mathrm{IC}$ and $K_\mathrm{IIC}$ fracture toughnesses which, as demonstrated in Sec. \ref{sec:calibration}, are controlled by the free parameters $\lambda_c$ and $\theta_c$, the critical stretch for a bond and the critical angle for three-body interactions, respectively.
Here we focus on the large-system-size, large-strain, and low-strain-rate limits with systems of size $L = 1600$ sheared to a strain of 1.0 at a rate of $10^{-5}$ to maximize the span of the power-law domain in $N(M)$.

First, we consider a fixed value of $\theta_c = 10^\circ$ and vary $\lambda_c$, which varies $K_\mathrm{IC}$ or the resistance to propagating an opening crack. 
As expected, decreasing the strength of bonds leads to more breakage as demonstrated in Fig. \ref{fig:render_lc}. 
In particular, at low $\lambda_c$ there are large regions of nearly fully pulverized material broken down into individual particles separating larger intact (although heavily damaged) grains. 
In the limit of very weak bonds, it is unsurprising that large swaths of the material break as vibrations emanating from cracks may have enough energy to break surrounding bonds.
As $\lambda_c$ increases, the width of these regions and the amount of visible damage in grains decrease.
It is possible that this effect could be counteracted by increasing the strength of damping in the system, as discussed at the end of Sec. \ref{sec:fr}.

\begin{figure}
\begin{centering}
	\includegraphics[width=0.40\textwidth]{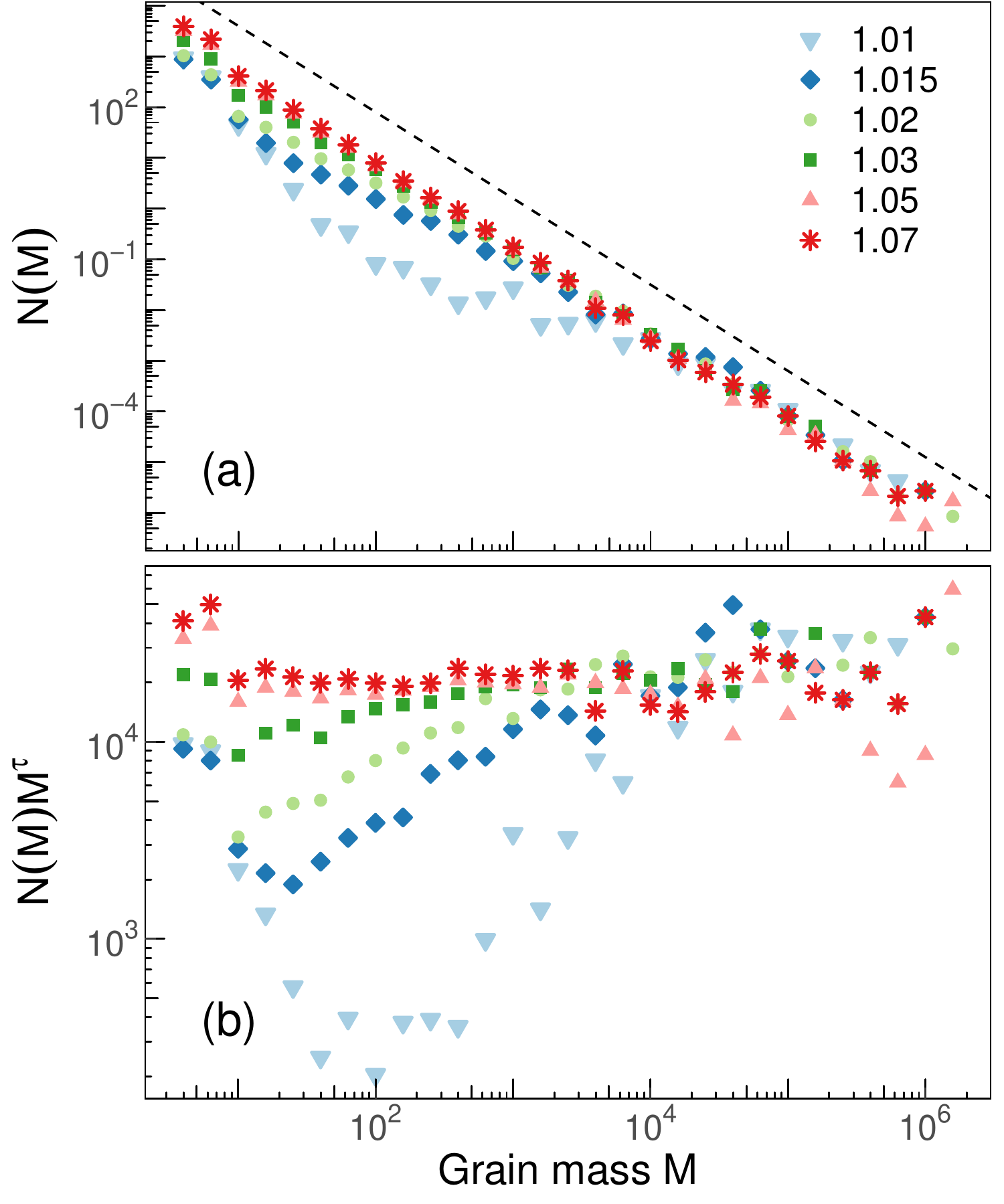}
	\caption{(a) Distributions $N(M)$ for a 2D system of size $L = 1600$, $k_a = 2.5$, and $\theta_c = 10^\circ$ sheared to a strain of 1.0 at a rate of $10^{-5}$ with the indicated values of $\lambda_c$. The dashed line represents a power law with exponent $\tau = 1.7$. (b) Distributions are normalized by $M^\tau$.} 
	\label{fig:NM_lc}
\end{centering}
\end{figure}

While there is markedly different behavior at small length scales, to determine whether the large-scale critical statistics depend on $\lambda_c$ we evaluate grain size distributions $N(M)$ (Fig.  \ref{fig:NM_lc}).
Although small grains with a mass of unity are not included in Fig. \ref{fig:render_lc}, there is a large excess at small $\lambda_c$, as expected.
In contrast, there is a deficit of grains with intermediate masses at small $\lambda_c$.
For instance, at $\lambda_c = 1.01$, grains with masses between $\sim 5$ and $10^4$ are underrepresented relative to systems with $\lambda_c \ge 1.05$.
The size of this domain and the magnitude of the under-representation shrinks with increasing $\lambda_c$ before converging around $\lambda_c = 1.05$. 
Despite this substantial change in the statistics of small and intermediate grains, no significant dependence on $\lambda_c$ is present in $N(M)$ at large $M$.
This suggests that $\lambda_c$ is irrelevant in the large-mass limit and does not change the fundamental critical nature of fragmentation.
Alternatively, $K_\mathrm{IC}$ only appears to affect the lower cutoff of the power-law behavior, $M_\mathrm{min}$, and the statistics of smaller grains.

\begin{figure}
\begin{centering}
	\includegraphics[width=0.40\textwidth]{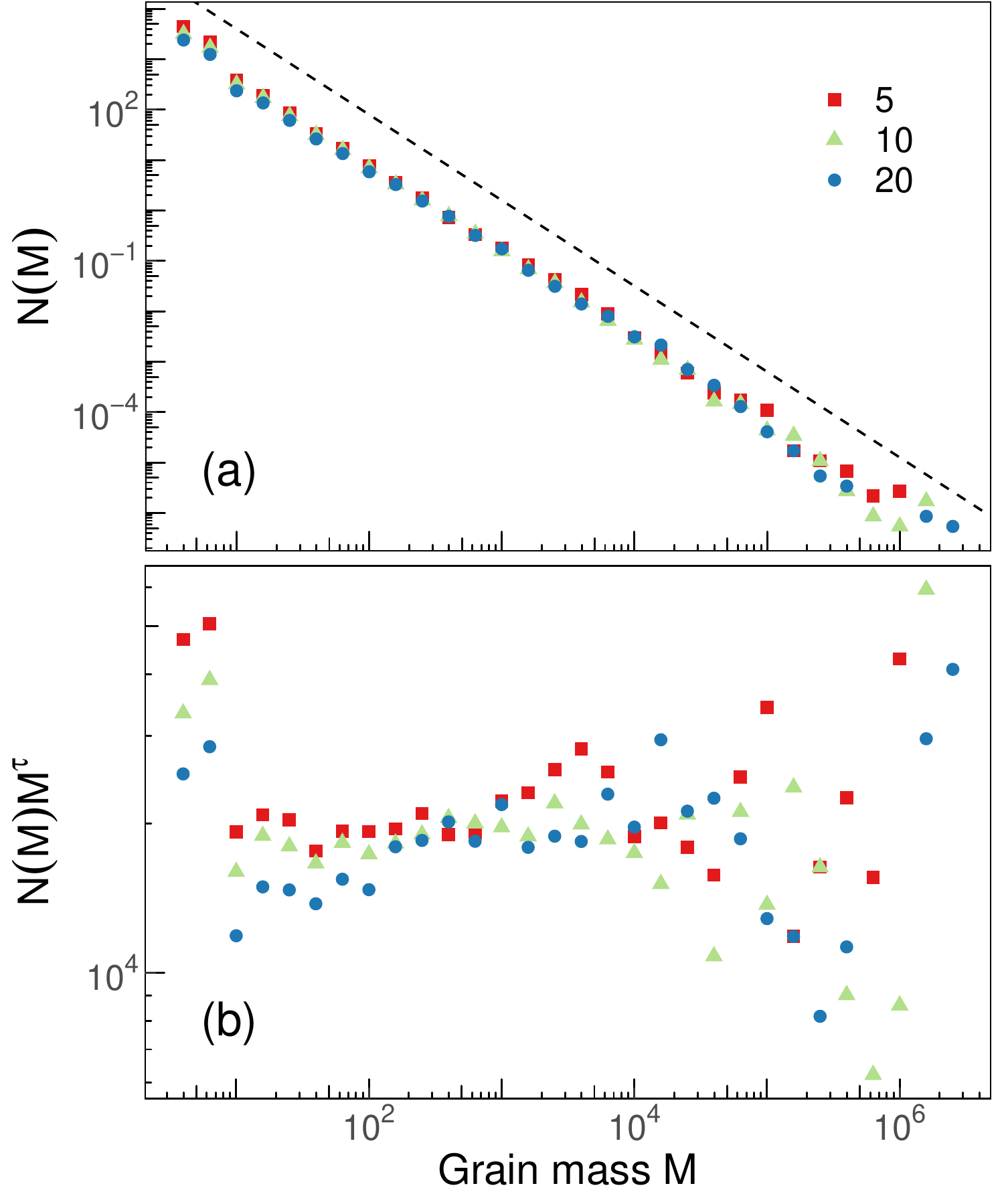}
	\caption{(a) Distributions $N(M)$ for a 2D system of size $L = 1600$, $k_a = 2.5$, and $\lambda_c = 1.05$ sheared to a strain of 1.0 at a rate of $10^{-5}$ with the indicated values of $\theta_c$ in units of degrees. The dashed line represents a power law with exponent $\tau = 1.7$. (b) Distributions are normalized by $M^\tau$.} 
	\label{fig:NM_tc}
\end{centering}
\end{figure}

Finally, we fix $\lambda_c = 1.05$ and vary $\theta_c$, which is equivalent to maintaining a constant mode I fracture toughness while varying the mode II fracture toughness $K_\mathrm{IIC}$ or the resistance to propagating a shear crack.
In Fig. \ref{fig:NM_tc}, $N(M)$ is remarkably robust to changes in $\theta_c$.
While there may be slight changes in the statistics of small grains, the magnitude of any potential differences is minimal and is largely masked by uncertainty in the data.
Therefore, similar to the above analyses, the critical behavior of fragmentation does not depend on the fracture toughness of the material in the limit of large masses.

\subsection{Rheology}
\label{sec:rheo}

Having explored the impact of strain, system size, strain rate, and material properties on fragmented grain size distributions in the above sections, we now zoom out and consider the macroscopic response of the system to shear.
In addition to studying the typical values of $k_a = 2.5$ (2D) and $3.0$ (3D), in this section we also include results with $k_a = 0$ or $\nu_\mathrm{PR} = 1/3$ in 2D and $1/4$ in 3D since this option limits computational costs (no three-body interactions) so simulations can reach slower strain rates. 
Focusing on the granular flow regime after fracture, or strains greater than 0.05 (Fig. \ref{fig:stress_strain}), we calculate the internal friction or stress ratio $\mu \equiv \sqrt{J_2}/P$, where $P$ is the mean pressure of the system to characterize its rheology.
For granular systems at low pressures in the hard-grain limit, rheology is often described in terms of an inertial number
\begin{equation}
I =  \dot{\epsilon} \langle d \rangle \sqrt{\rho/P} \ \ ,
\end{equation}
where $\langle d \rangle$ is the average grain diameter and $\rho$ is the density of the system \cite{Jop2006}.
In a $\mu(I)$ model, $\mu$ grows as a function of increasing $I$, as demonstrated in many works \cite{Peyneau2008,DeCoulomb2017,DeGiuli2015,Srivastava2021,Clemmer2021c} including DEM studies of aspherical \cite{Salerno2018} and polydisperse grains \cite{Yohannes2010, Gu2016}.

\begin{figure}
\begin{centering}
	\includegraphics[width=0.40\textwidth]{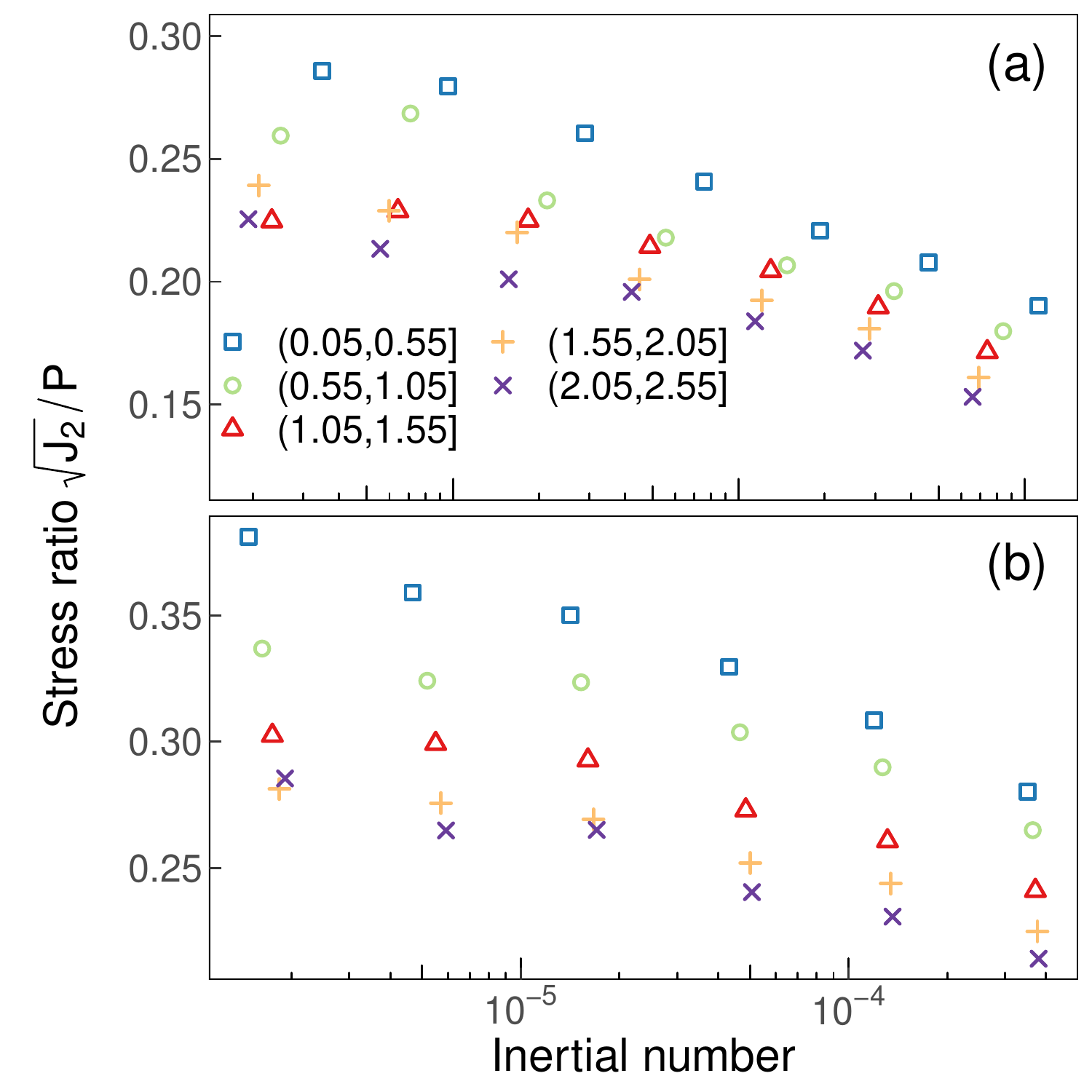}
	\caption{Stress ratio $\mu$ as a function of the inertial number $I$ for systems of size (a) $L = 1600$ in 2D and (b) $L = 200$ in 3D at $k_a = 0.0$ (solid lines) and $2.5$ in 2D and $3.0$ in 3D (dashed lines). Data is averaged over the indicated strain intervals.} 
	\label{fig:mu_of_I}
\end{centering}
\end{figure}

In contrast to traditional studies of granular rheology, the system in this article uniquely does not have a fixed average grain size $\langle d \rangle$.
Instead, $\langle d \rangle$ depends on both strain and rate.
Averaging over intervals of 0.5 (50\%) strain, both the stress ratio $\mu$ and the inertial number $I$ are calculated in large systems at different strain rates, extending down to $\dot{\epsilon} = 10^{-6}$ in 2D and $3 \times 10^{-6}$ in 3D at $k_a = 0$.
Results are plotted in Fig. \ref{fig:mu_of_I}. 
Here $\mu$ interestingly decreases with increasing inertial number (and strain rate) as well as increasing strain.
This reduction fundamentally reflects the fact that the system fragments into smaller grains at higher rates (Fig. \ref{fig:render_rate}) and that more of the system has broken up at larger strains (Fig. \ref{fig:render_strain}).
Note that in these simulations, the pressure is not held constant such that $I$ is calculated using an average pressure $P$ which evolves with strain and rate and may not reflect the hard-grain limit typically explored in DEM simulations of granular flow.
However, a confining pressure is always maintained postfailure.

As observed in Fig. \ref{fig:stress_strain}, the shear stress does actually increase with strain rate; however, the pressure increases at a faster rate, leading to the decrease in the stress ratio seen here.
In future work, it would be valuable to explore fragmentation under alternate loading conditions such as constant pressure instead of constant volume. 
In particular, varying the confining pressure could test the impact of cavity formation on fragmentation.
Additionally, experiments of granular breakage under shear by \citet{Xu2018} found a potential pressure-dependence on the power-law exponent $\tau$.

At larger $k_a$ or lower $\nu_\mathrm{PR}$, stress ratios are higher as curves shift upward.
The difference is around $0.015$ in both 2D and 3D.
This effect likely originates from changes in the aspherical shape of grains seen in Fig. \ref{fig:render_k} as increasing asphericity of grains is associated with greater stress ratios in flow \cite{Salerno2018}, but could also emerge from the slight differences in polydispersity seen in Fig. \ref{fig:NM_k}.

One particularly intriguing feature of the rheology in Fig. \ref{fig:mu_of_I} is that $\mu$ appears to decay logarithmically with increasing inertial number.
Similar behavior is seen when $\mu$ is plotted as a function of strain rate. 
Such logarithmic velocity weakening is often an important feature in rate and state friction models \cite{Daub2010} and this demonstrates that it can naturally emerge from fragmentation in sheared granular materials.
Furthermore, these findings highlight how the distribution of grain sizes can have a significant influence on rheology, emphasizing the need to explore rheology beyond the low-pressure, monodisperse, and spherical limits which do not always reflect granular material found in nature.
For instance, it is unclear whether the factor of $\langle d \rangle$ in $\mu(I)$ rheology is an appropriate metric for highly polydisperse granular material such as the power-law distributed set of grains seen here.
Potentially another metric, such as $\sqrt{\langle d^2 \rangle}$, would be more informative.
This underlines a need to understand granular rheology beyond narrow size distributions.

Furthermore, this data prompts questions about the nature of the crossover between the quasistatic and finite-rate limits. 
In contrast to the granular structure, quantified in terms of the mean-square grain size in Fig. \ref{fig:msq_collapse}, there generally is no clear saturation in the strength ratio with decreasing strain rate, despite extending to slower rates in a similarly sized system at $k_a = 0.0$.
While some data series may exhibit this saturation (such as strains between 1.05 and 1.55), this may just reflect the uncertainty in the measured stress ratios.
Therefore, it is likely that the rheology actually crosses over to the quasistatic limit at a lower rates than the grain size distribution and is controlled by a separate mechanism.
In smaller systems, there is a clear saturation in $\mu$ with decreasing strain rate; however, a full finite-size analysis of the rheology is beyond the scope of this paper.

\section{Summary}
\label{sec:summary}

This article has provided a systematic exploration of simulated fragmentation in isotropic, brittle solids under shear, particularly focusing on the impact of material properties on the scaling of grain size distributions in 2D.
Using a simple bonded particle model with breakable three-body angular interactions, we were able to individually control Poisson's ratio and mode I and mode II fracture toughnesses which are calibrated using only three free parameters.
While the rate of production of new grains and some noncritical aspects of fragmentation were found to depend on material properties, all systems reached the same critical power-law distribution of grain sizes in the quasistatic large-strain limit. 
Universal behavior in fragmentation has been proposed and tested several times in the literature \cite{Oddershede1993, Inaoka1996, Astrom2000, Astrom2004, Einav2007, Timar2010} and this work provided a comprehensive extension and validation of this idea through the breadth of material properties studied under a wide range of conditions in 2D and by considering the less-studied loading geometry of shear flow.
Furthermore, using a finite-size and finite-rate scaling theory proposed in our earlier work, we demonstrated that several additional critical exponents, summarized in Table \ref{table:exponents}, were consistent across a wide range of elastic properties.
A more limited exploration was also performed in 3D where we found similar quasistatic and finite-rate behavior with the addition of three-body interactions, although we did not systematically explore finite-size effects or test a wide range of material properties.
Finally, we identified several other unique features of a fragmenting material under shear flow including an abnormal rheology where the stress ratio decays with increasing rate.

As systems fracture and fragment in shear at low strain rates, the number of grains of a given mass depended significantly on Poisson's ratio $\nu_\mathrm{PR}$ as more grains were generated immediately upon fracture at small $\nu_\mathrm{PR}$ but the rate of subsequent increase was slower in comparison to systems with large $\nu_\mathrm{PR}$ (Sec. \ref{sec:strain}).
It is possible that the growth in the number of grains with strain is described by a power law with an exponent $\phi$.
If so, $\phi$ might decrease from $\sim0.55$ at $\nu_\mathrm{PR} = 0.25$ to $\sim 0.25$ at $\nu_\mathrm{PR} = -0.11$.
However, this dependence on $\nu_\mathrm{PR}$ appeared to be the exception as the distribution of grain sizes otherwise demonstrated minimal dependence on material properties.
Shortly after yield, a power-law domain with an exponent $\tau$ was identified in $N(M)$.
This domain subsequently grew with strain until a strain of $\approx 1.0$ (or 100\%).
This saturation at large strains reflects results from experimental studies of comminution in shear \cite{Marone1989, Coop2004}.
Although there appeared to be a slight increase in the exponent $\tau$ with decreasing $\nu_\mathrm{PR}$ of about $10\%$, this was less than uncertainty in measurements and could simply be due to slightly different manifestations of finite-rate or finite-size effects and therefore there was no detectable $\nu_\mathrm{PR}$-dependence on $\tau$.
Similarly, varying the fracture toughnesses in Sec. \ref{sec:kc} revealed no detectable impact on $\tau$, although the onset of the power-law domain shifts to larger masses with decreasing mode I fracture toughness due to a tendency for the material to produce an excess of small grains and a dearth of intermediate-size grains.

In the quasistatic limit, a scaling relation for finite-size effects in $N(M)$ was described in Eq. \eqref{eq:nqs} based on two nontrivial exponents $\gamma$ and $\alpha$ and tested in Sec. \ref{sec:size}.
For the first exponent, the number of grains of a given mass $M$ grew as a power of the linear system size $L$ with an exponent $\gamma$ less than $d = 2$ as demonstrated by directly scaling $N(M)$ by $L^\gamma$.
For the second exponent, the mass of grains at the upper cutoff of the power-law regime $M_\mathrm{cut}$ was found to grow as a power of $L$ as $M_\mathrm{cut} \sim L^\alpha$ where again $\alpha < 2$, implying both the number of grains and the size of the largest grain grow subextensively.
By calculating the average grain size, we derived a scaling relation in Eq. \eqref{eq:scaling_relation} between these exponents and $\tau$ based on conservation of mass.
Variations in $\nu_\mathrm{PR}$ had no detectable effect on any of these exponents or scalings.

Beyond the quasistatic limit, this paper also presented a description of fragmentation at high rates in Sec. \ref{sec:fr}.
With increasing rate, fragmentation produced a finer set of grains and the distribution $N(M)$ unusually became shallower, possibly reflecting a changing power-law exponent.
While we could not fully determine the origin of this effect, it is possible that this is simply an instance of noncritical behavior at particularly high rates that may become irrelevant in the limit of infinitely small strain rates and infinitely large system sizes. 
However, this effect is still an important topic for further study, especially as it may relate to variations in measured values of $\tau$ in real-world materials and experiments \cite{Turcotte1986}.

By combining studies of finite sizes and rates in Sec. \ref{sec:frs}, we observed that fragmentation only depends on either the size of the system or the rate, but not both. 
We proposed that there exists a diverging length scale $\xi \sim \dot{\epsilon}^{-\nu}$, where $\nu$ is another exponent less than unity, and that the scaling of $N(M)$ only depends on the smallest length scale, either $\xi$ or $L$. 
In the finite-rate limit where $\xi < L$, a scaling ansatz for $N(M)$ was derived in Eq.~\eqref{eq:nfr}.
Combining the quasistatic and finite-rate expressions for $N(M)$, we then constructed a finite-size scaling relation for the second moment in Eq.~\eqref{eq:ansatz_m2} which collapsed data across system size and rate, yielding an estimate of $\nu$. 
This collapse was consistent across values of Poisson's ratio $\nu_\mathrm{PR}$.

Finally, we considered the system's rheology in Sec. \ref{sec:rheo} as a unique example of a granular system with evolving polydispersity.
As grains begin to flow after fracture, the internal friction decays as the logarithm of increasing strain rate.
Such logarithmic weakening is an often studied topic in geophysics and friction, and this work demonstrated how fragmentation can emergently lead to this important behavior.
This emphasizes a need for further studies of flowing polydisperse granular materials with different grain size distributions to characterize changes rheology.

In conclusion, we find the observed critical behavior is remarkably robust to changes in material properties with no detectable deviation in exponents, summarized in Table \ref{table:exponents}, with the possible exception of $\phi$, which cannot be concluded to be a real critical exponent due to a limited scaling domain.
However, there are still countless unanswered questions about the physics of fragmentation.
For instance, additional studies extending analysis of finite-size effects to three-dimensional systems are needed and variations in the damping strength were not investigated.
Alternative loading geometries, in particular stress-controlled deformations at various mean pressures, need to be studied.
Low damping strengths could introduce interesting inertial effects which have been found to affect avalanches in the yielding transitions \cite{Salerno2012, Salerno2013, Karimi2017} and the role of energy dissipation needs to be better understood.
In comparison to results from our recent paper \cite{Clemmer2022}, the results in this text were limited to smaller system sizes and higher strain rates due to the additional computational complexity of controlling Poisson's ratio.
This highlights an ongoing need to develop more efficient bond models which can control elastic properties, in particular efficient methods that can model higher Poisson's ratios which were inaccessible to this work.

\begin{acknowledgments}
The authors thank Joseph Monti for useful discussions.
Calculations were performed at the Maryland Advanced Research Computing Center, the DoD High Performance Computing Modernization Program, and at Sandia National Laboratories. Research was sponsored by the Army Research Laboratory and was accomplished under Cooperative Agreement Number W911NF-12-2-0022. The views and conclusions contained in this document are those of the authors and should not be interpreted as representing the official policies, either expressed or implied, of the Army Research Laboratory or the U.S. Government. The U.S. Government is authorized to reproduce and distribute reprints for Government purposes notwithstanding any copyright notation herein.

This article has been authored by an employee of National Technology \& Engineering Solutions of Sandia, LLC under Contract No. DE-NA0003525 with the U.S. Department of Energy (DOE). The employee owns all right, title and interest in and to the article and is solely responsible for its contents. The United States Government retains and the publisher, by accepting the article for publication, acknowledges that the United States Government retains a non-exclusive, paid-up, irrevocable, world-wide license to publish or reproduce the published form of this article or allow others to do so, for United States Government purposes. The DOE will provide public access to these results of federally sponsored research in accordance with the DOE Public Access Plan https://www.energy.gov/downloads/doe-public-access-plan.
This paper describes objective technical results and analysis. Any subjective views or opinions that might be expressed in the paper do not necessarily represent the views of the U.S. Department of Energy or the United States Government.
\end{acknowledgments}

\bibliographystyle{apsrev4-1}

\end{document}